\documentclass[a4paper, 12pt]{kluwer}
\usepackage[left = 3.5cm, right = 3.5cm, top = 3.5cm, bottom = 3.5cm]{geometry}
\usepackage{import}
\usepackage{adjustbox}
\usepackage{graphicx}
\usepackage{svg}
\usepackage{multirow}
\usepackage{booktabs}
\usepackage[utf8]{inputenc} 
\usepackage[hyphens]{url}

\usepackage{color,colortbl}

\definecolor{Gray}{gray}{0.95}
\definecolor{LightCyan}{rgb}{0.88,1,1} 
\definecolor{LightCyan}{RGB}{206,255,255}
\definecolor{Yellow}{RGB}{252, 243, 207}

\usepackage{array}
\usepackage{fancyhdr}
\usepackage{subfig}
\usepackage{verbatim}

\usepackage{color,soulutf8}

\definecolor{ChangesColor}{RGB}{252, 243, 207}
\definecolor{celeste}{RGB}{252, 243, 207}
\definecolor{celeste2}{RGB}{204, 238, 255}
\definecolor{white}{RGB}{255,255,255}

\definecolor{rojo}{RGB}{255,204,204}

\DeclareRobustCommand{\hlreview}[1]{{\sethlcolor{white}\hl{#1}}}

\DeclareRobustCommand{\celeste}[1]{{\sethlcolor{white}\hl{#1}}}

\DeclareRobustCommand{\nuevo}[1]{{\sethlcolor{white}\hl{#1}}}

\DeclareRobustCommand{\final}[1]{{\sethlcolor{white}\hl{#1}}}
\DeclareRobustCommand{\revisado}[1]{{\sethlcolor{white}\hl{#1}}}

\usepackage{navigator}
\usepackage{tikz,xcolor}

\definecolor{lime}{HTML}{A6CE39}
\DeclareRobustCommand{\orcidicon}{
	\begin{tikzpicture}
	\draw[lime, fill=lime] (0,0) 
	circle [radius=0.16] 
	node[white] {{\fontfamily{qag}\selectfont \tiny ID}};
	\draw[white, fill=white] (-0.0775,0.1) 
	circle [radius=0.005];
	\end{tikzpicture}
	\hspace{-3mm}
}
 
\foreach \x in {A, ..., Z}{%
	\expandafter\xdef\csname orcid\x\endcsname{\noexpand\urllink{https://orcid.org/\csname orcidauthor\x\endcsname}{\noexpand\orcidicon}}

}

\fancypagestyle{firststyle}
{
   \fancyhf{}
   \fancyfoot{}
}
\begin{document}     
\thispagestyle{firststyle}
\begin{opening}         
\title{\textbf{Regional Differences in Information Privacy Concerns After the Facebook-Cambridge Analytica Data Scandal\footnote{This version corresponds to the accepted manuscript (29 November 2021). For the final published version
and citation please see \url{https://doi.org/10.1007/s10606-021-09422-3}}}}

\thispagestyle{firststyle}

\author{Felipe González-Pizarro$^{1^\dagger}$\orcidB{}, Andrea Figueroa$^{2}$, Claudia López$^{1}$\textsuperscript{*}\orcidA{} \& Cecilia Aragon$^{2}$}
\institute{$^{1}$Departamento de Informática, Universidad Técnica Federico Santa María, Chile (felipe.gonzalezp.12@sansano.usm.cl, claudia@inf.utfsm.cl) $^{2}$Human-Centered Data Science Lab, University of Washington, Seattle, United States (afigue@uw.edu, aragon@uw.edu) $^{\dagger}$Department of Computer Science,  University of British Columbia, Vancouver, Canada (felipegp@cs.ubc.ca)} 

\runningauthor{Felipe González-Pizarro et al.}
\runningtitle{Regional Differences in Information Privacy Concerns...}

\begin{abstract}

\noindent

\hlreview{
While there is increasing global attention to data privacy, most of their current theoretical understanding  is based on research conducted in a few countries. Prior work argues that people's cultural backgrounds might shape their privacy concerns; thus, we could expect people from different world regions to conceptualize them in diverse ways. We collected and analyzed a large-scale dataset of tweets about the \#CambridgeAnalytica scandal in Spanish and English to start exploring this hypothesis. We employed word embeddings and qualitative analysis to identify which information privacy concerns are present and characterize language and regional differences in emphasis on these concerns. Our results suggest that related concepts, such as regulations, can be added to current information privacy frameworks. We also observe a greater emphasis on data collection in English than in Spanish. Additionally, data from North America exhibits a narrower focus on awareness compared to other regions under study.  Our results call for more diverse sources of data and nuanced analysis of data privacy concerns around the globe.}

\end{abstract}


\end{opening}           
\thispagestyle{firststyle}

\section{Introduction }

The right to control one's personal information has gained significant importance lately \cite{lee2019information}. Indeed, 58\% of the countries have data protection and privacy legislation, while another 10\% have drafted legislation about it \cite{unctd}. This broad interest is related to the massive amount of personal data collected by information systems and the risk that such information could be wrongly distributed online \cite{lee2019information}. 

The study of information privacy has advanced our understanding of individuals' concerns regarding organizational practices associated with collecting and using their personal information \cite{smith1996information}. \nuevo{However, a literature review revealed a strong bias towards USA-centered studies across privacy concerns literature and warned about the limitations to generalizability this entails} \cite{belanger2011privacy,OKAZAKI2020458}. The review's authors hypothesized that individuals from different world regions have diverse cultures, values, and laws, which can, in turn, result in different conceptualizations of information privacy and its impacts \cite{belanger2011privacy,MOHAMMED2017254}. To study these differences, privacy research has often relied on survey-based studies \cite{cockcroft2016relationship}. For example, a questionnaire was applied to explore differences in privacy perceptions between Facebook users from Germany and the USA \cite{krasnova2010privacy}, and a cross-national survey was conducted to evaluate information attitudes of consumers in the USA and Brazil \cite{markos2017information}. These multi-country privacy studies have had limited sample sizes, which makes the results difficult to generalize \cite{lee2019information,huang2016privacy}. They also tend to be focused on one or two cultures, usually including the USA \cite{cockcroft2016relationship}. Hence, multi-country information privacy research is still needed to extend our understanding of this increasingly relevant topic around the globe \cite{adu2019individuals,va_zou2018ve}.

We propose an alternative approach to study information privacy concerns over a large geographical scope. This work combines word embeddings, open coding, and content analysis to examine tweets related to a large data breach scandal. We seek to characterize similarities and differences in privacy 
terms across people who tweet about this issue in different languages and from different world regions. Inspired by \cite{rho2018fostering}, where text analysis was used to analyze answers about individuals' privacy concerns, we analyze the semantic context in which privacy-related terms were used in tweets written by different groups of people. 

We focus on the Facebook–Cambridge Analytica data scandal. \nuevo{In 2018, the firm Cambridge Analytica was accused of collecting and using the personal information of more than 87 million Facebook users without their authorization}  \cite{venturini2019api,isaak2018user,lapaire2018content}. The scandal sparked multiple conversations over technology's societal impact and risks to citizens' privacy and well-being worldwide. Opinions, facts, and stories related to it took place \nuevo{on} different social media platforms such as Twitter, where the hashtag \#DeleteFacebook became a trending topic for several days \cite{lin2018deletefacebook,mirchandani2018delete}. 

\hlreview{We analyze more than a million public tweets in Spanish or English that use hashtags or keywords related to the scandal. We divide the dataset by language (Spanish and English) and regions (Latin America, Europe, North America, and Asia) and create word-embeddings for each subset. Then, we systematically analyze and compare the semantic context of four keywords, such as \textit{data}, \textit{privacy}, \textit{user}, and \textit{company}, across the embeddings. We contrast our results with one of the most used information privacy concerns framework to find terms and tweets matching different concerns. Then, we test a null hypothesis that there is no difference in emphasis on information privacy terms across languages and world regions. In this process, we discover the presence of related concepts that could be integrated into information privacy frameworks, such as regulations. We also observe statistically significant language differences in emphasis on data collection and significant regional differences in emphasis on awareness. Finally, we discuss the implications of our results.}

We summarize prior work on information privacy concerns in Section \ref{secinformationprivacydifferences}. Section \ref{sec:research_question} introduces our research question \hlreview{and hypothesis}. Section \ref{sec:data_and_methods} details our research method, while Section \ref{sec:results} reports on our findings. Section \ref{sec:discussion} offers a discussion of our results, limitations, and future work. Finally, Section \ref{sec:conclusion} provides our conclusions.


\section{Information privacy concerns}\label{secinformationprivacydifferences}

\nuevo{Information privacy concerns emerge when an individual ``feels threatened by a perceived unfair loss of control over their privacy by an information-collecting body''}\cite{lee2015compensation}. \nuevo{Previous research argues that information privacy concerns are a multidimensional construct}\cite{JOZANI2020106260,correia2017information,HERAVI2018441}. \hlreview{A multidimensional approach allows identifying to what extent users are concerned about different aspects of information privacy} \cite{YUN2019570,10.2307/43825946,10.2307/43825946}.
\hlreview{Different authors have proposed alternative 
conceptualizations to measure information privacy concerns. 
We briefly review the most adopted ones in the following subsection. Then, we summarize prior work on differences in privacy concerns across countries, regions, and other characteristics.}

\celeste{Prior research has examined privacy concerns from other perspectives as well }\cite{HERAVI2018441,gerber2018explaining,YUN2019570}. 
\celeste{A vast portion of privacy research on social networking sites has focused on examining users' privacy behaviors, such as the intention to provide personal information or 
transact online }\cite{WISNIEWSKI201795,kokolakis2017privacy,gerber2018explaining,OGHAZI2020531,su12198286,HERAVI2018441,markos2017information}. \nuevo{In a similar direction, several studies have 
investigated the use of privacy setting configurations} \cite{va_10.1145/2675133.2675256,IJoC3208,WISNIEWSKI201795,doi:10.1080/0144929X.2020.1831608}. 
\nuevo{Rather than centering on behavior or behavioral intention, we focus our review on information privacy concerns 
that characterize general personal dispositions }\cite{c2013empirically}.\nuevo{ We think that this part of the literature 
aligns better with what people can say, in a declarative way, about data privacy on social media
.}

\subsection{Assessing information privacy concerns}

Two questionnaires have been widely used to evaluate individuals' information privacy concerns \cite{belanger2011privacy,cockcroft2016relationship,morton2014desperately}: Concerns for Information Privacy (CFIP) and Internet Users' Information Privacy Concerns (IUIPC). 

\subsubsection{CFIP: Concerns for Information Privacy}

The CFIP framework \cite{smith1996information} 
focuses on  individuals' perceptions of how organizations use and protect personal information \cite{van2006concern}. 
CFIP identifies four 
dimensions:
 
\begin{itemize}
\item \textit{Collection:} concerns about personal data that is collected over time; 
\item \textit{Unauthorized secondary use:} concerns about organizations using personal data for another purpose without the individual's authorization; 
\item \textit{Improper access:} concerns about unauthorized people having access to personal data;  \item \textit{Errors:} 
concerns about adequate protections from deliberate and accidental errors in personal data. 
\end{itemize}

To measure them, \inlinecite{smith1996information} proposed and validated a 15-item questionnaire. The CFIP questionnaire was validated by surveying 355 consumers from the USA and applying confirmatory factor analysis (CFA) \cite{stewart2002empirical}. So far, this questionnaire had been considered as one of the most established methods to measure quantitatively information privacy concerns \cite{harborth2018german} and had been widely used in the literature 
\cite{harborth2017privacy,stewart2002empirical}. 
However, the CFIP and its measurement instrument were originally defined for users in an offline context \cite{palos2017behavioral}. As the Internet enabled new ways to collect and process data, it was expected that new concerns about information privacy might emerge \cite{malhotra2004internet}, and a new framework was proposed: the IUIPC. 

\subsubsection{IUIPC: Internet Users' Information Privacy Concerns}
\label{iuipc}


\inlinecite{malhotra2004internet} introduced the IUIPC framework and conceptualized Internet users' concerns about information privacy from a perspective of fairness. Drawing from social contract theory \cite{c2013empirically}, \inlinecite{malhotra2004internet} argue that personal data collection is perceived to be fair when a user has control over their personal data and is informed about the intentions that organizations have about how to use it. The IUIPC includes three constructs:

\begin{itemize}
    \item \textit{Collection:} concerns about the amount of personal data owned by others compared to the perceived benefits \cite{malhotra2004internet}. 
    It is related to the perceived fairness of the outcomes one receives. 
    Users provide information if they expect to obtain something of value after a cost-benefit analysis of a transaction. 
      
    \item \textit{Control:} concerns about control over personal information, including approval, modification of collected data, and opportunity to opt-in or opt-out from data collection \cite{malhotra2004internet}. 
    It is related to the perceived fairness of the procedures that maintain personal data. 
    \item \textit{Awareness:} concerns about personal awareness of organizational information practices \cite{malhotra2004internet}. It relates to issues of transparency of the procedures and specificity of information to be used.  

\end{itemize}



A 10-item questionnaire to assess these constructs was validated in \cite{malhotra2004internet}. The questionnaire has been widely used to this day \cite{YUN2019570,raber2018privacy} because it considers the Internet context, and it can explain more variance in a person's willingness to transact than CFIP \cite{rowan2014observed}. Recent work has explored text mining as an alternative research method to identify IUIPC dimensions. \inlinecite{raber2018privacy} found that IUIPC dimensions can be derived from written text. They observed a correlation between IUIPC concerns, as measured by the questionnaire, and LIWC language features of social media posts from a sample of 100 users.

\subsubsection{Other instruments of assessment}

The Westin-Harris Privacy Segmentation Index 
measures individuals' attitudes and concerns about privacy and how they vary over time \cite{kumaraguru2005privacy} based on answers to three questions \cite{egelman2015predicting,woodruff2014would}. It categorizes individuals into three groups 
\cite{kumaraguru2005privacy,da2018information,motiwalla2014privacy}:
   \textit{Fundamentalists} are highly concerned about sharing their data, protect their personal information, 
   prefer privacy controls over consumer-service benefits, and are in favor of new privacy regulations;
\textit{Pragmatists} tend to seek a balance between the advantages and disadvantages of sharing personal information before arriving at a decision;
\textit{Unconcerned} users believe there is a greater benefit to be derived from sharing their personal information, trust organizations that collect their personal data and 
are the least protective of their privacy. 



The Westin-Harris' index was introduced as a way to meaningfully classify internet users based on their attitude toward privacy and their motivations to disclose personal information 
\cite{torabi2016towards}. 
It 
has been used for several decades. However, recent studies have raised questions about its validity \cite{egelman2015predicting}. Prior work has failed to establish a significant correlation between the Westin-Harris' segmentation and context-specific, privacy-related actual or intended behaviors 
\cite{consolvo2005location,woodruff2014would,egelman2015scaling}. 



The existence of a mismatch between privacy concerns and 
privacy behaviors, known as the 
``privacy paradox'' \cite{kokolakis2017privacy,dienlin2015privacy}, motivated the creation of a new measurement instrument. Buchanan's Privacy Concern scale aims to capture different aspects of the paradox. \inlinecite{buchanan2007development} developed three privacy scales: two of them assess privacy behavior, and the third one measures information privacy concerns. 
%
%
However, some limitations have been identified. 
\nuevo{Their scales are not able to identify different privacy dimensions, but only one, which appears to map onto the general concept of privacy concern.}
Thus, a more fine-grained examination is desirable to improve the design of this scale \cite{buchanan2007development}.

Because our study focuses on people's comments about a specific information privacy scandal (and not their privacy behavior), our work will mostly build upon the information privacy concerns frameworks, particularly the IUIPC. 

\subsection{Differences on information privacy concerns}

Information privacy concerns 
can vary across individuals based on peoples' perceptions and values \cite{buchanan2007development}. People may have different concerns even if they experience the same situation \cite{lee2015compensation}. It has been argued that information privacy concerns can be influenced by different factors 
\cite{smith2011information}, \hlreview{such as} 
%
national culture \cite{cho2009multinational,huang2016privacy,cao2008user,krasnova2010privacy}, and 
individuals' demographics (e.g., age, gender)  \cite{zukowski2007examining,lee2019information,jai2016privacy,rowan2014observed,cho2009multinational,markos2017information}. 
We review these factors below.
\subsubsection{National Culture}

While 
there are similarities in what privacy means 
across cultures \cite{cockcroft2016relationship}, there is no universal consensus on its definition \cite{cannataci2009privacy}. 
According to \inlinecite{newell1995perspectives}, several cultures 
do not possess an equivalent term to the English' privacy definition in their own language, e.g., Arabic, Dutch, Japanese, and Russian. 
Nevertheless, this does not mean that these cultures 
lack a sense of privacy \cite{newell1995perspectives}. 
Every society appreciates 
privacy in some way, but the expression of it varies 
\cite{cho2009multinational}. 

The concept of national culture 
has been studied as one of the 
factors related to information privacy concerns \cite{nov2009social,malhotra2004internet,bellman2004international}. National culture can be defined as ``the collective mindset distinguishing the member of one nation from another''\cite{cho2009multinational}.
%
%
Hofstede's cultural dimensions theory \cite{hofstede1983national} 
has been the most used conceptual model to study cultural differences in this context. This trend is expected since Hofstede's theory has been widely used to study the relationship between culture and technology \cite{leidner2006review}, even though there are a number of criticisms of this theory \cite{terlutter2006globe}. 
The latest version of this theory proposes six cultural dimensions \cite{hofstede2011dimensionalizing}. Among them, the \textit{individualism/collectivism} dimension has been found relevant to information privacy concerns. %
 \textit{Individualism/collectivism} refers to the extent to which individuals are part of groups beyond their immediate families. 
Differences \celeste{in} information privacy concerns have been explained using some 
cultural dimensions at a country and regional level (see Table \ref{tab:summary_work_related_culture}). 
Participants from \hlreview{individualistic} 
countries (Australia and United States) exhibited a higher level of online privacy concerns than individuals from collectivist countries \cite{cho2009multinational}. The authors' rationale is that high individualism is associated with an emphasis on private life and independence from the collective; thus, people from individualist countries are more worried about privacy intrusions. 
In the same direction, \inlinecite{bellman2004international} 
found that controlling for internet experience and privacy regulations, 
people from countries with high individualism show \celeste{deeper} 
concern about two CFIP dimensions: 
\textit{unauthorized secondary use} and \textit{improper access}. 


On the other hand, no regional differences in privacy concerns were found through 
online surveys with 226 English-fluent crowd workers 
from six regions (Africa, Asia, Western Europe, Eastern Europe, North America, and Latin America). 
The authors argued that it is unclear if their finding is due to true similarities or \hlreview{a lack of enough} power 
in measuring privacy concerns 
\cite{huang2016privacy}. 

\begin{table}[ht]
\caption{Culture and information privacy concerns}

\label{tab:summary_work_related_culture}
\begin{tabular}{p{0.12\linewidth}p{0.35\linewidth}p{0.17\linewidth}p{0.25\linewidth}}
\hline
Independent variables & 
Method & \# participants and origin & Key findings \\ \hline
National culture & 5-item questionnaire \cite{cho2009multinational}, based on a unidimensional conceptualization of online privacy concerns. Items were comprehensive enough to measure  general concerns about online privacy. 
& 1261 from Seoul, Singapore, Bangalore, Sydney, New York & 
Participants from \hlreview{individualistic} countries 
exhibited higher 
concern about online privacy \cite{cho2009multinational}
\\
National culture & 

15-item questionnaire (CFIP) \cite{smith1996information}, based on a multidimensional constructive model of privacy concerns (collection, unauthorized secondary use, improper access, errors). 
& 534 from 38 countries 
&  Participants from \hlreview{individualistic} countries 
showed higher concern about improper access and 
secondary use 
\cite{bellman2004international} 
\\
Regional culture & 
4-item questionnaire \cite{dinev2006extended}, based on a unidimensional conceptualization of privacy concerns, which is defined as apprehension about how online personal information is used by others. 
& 226 from Africa, Asia, Western and Eastern Europe, North and Latin America & No regional differences in privacy concerns were found \cite{huang2016privacy}\\ \hline
\end{tabular}
\end{table}


\subsubsection{Language
} 

Relatedly, the Sapir-Whorf hypothesis suggests that the structure of anyone's native language influences the world-views they will acquire 
\cite{kay1984sapir}. \hlreview{Depending on the language, a message is coded and decoded differently based on standardized language norms and culture \mbox{\cite{zarifis2019exploring}}}. Thus, individuals who speak different native languages could think, perceive reality and organize the world around them in different ways \cite{hussein2012sapir}. 

Previous work has explored how user-generated content can reveal different views about the same issues among people who write in different languages. 
\inlinecite{jiang2017mapping} conducted a semantic network analysis to examine the semantic 
differences that emerge 
from the Wikipedia articles about China. 
Results suggest that Chinese-speaking 
and English-speaking contributors framed articles about China in different and even opposite ways, which were aligned to their 
national cultures and values. 
The Chinese version 
framed them from perspectives of authority respect, emphasizing harmony and patriotism. 
Articles in English were written from \celeste{the} point of view that is distinctive of many Western societies: the core value of democracy. 


A potential role of the spoken language in the information privacy context has also been studied. \inlinecite{li2017cross} created a cross-cultural privacy prediction model. 
The model applies supervised machine learning to predict users' decisions on the collection of their personal data. Using answers from an online survey 
of 9,625 individuals from 8 countries on four continents: Canada, China, Germany, United States, United Kingdom, Sweden, Australia and, India,they found that the model's prediction accuracy 
improved  \nuevo{when adding individual's language (English, Chinese, French, Swedish, and German)  
or Hofstede's cultural dimensions}. 
Our work will build upon this line of reasoning to \hlreview{deepen} our understanding of information privacy concerns across the globe.

\subsubsection{Other individual characteristics}

Even though our work will not address the relationship between demographics and information privacy concerns, we will briefly review the literature about this topic.

Prior studies suggest that older Internet users are more concerned about online information privacy than younger ones \cite{cho2009multinational}. 
Older participants were more sensitive to privacy issues and exhibited a greater desire to control the amount of information collected about them \cite{zukowski2007examining}. In contrast, 
younger users declared themselves to be more willing to share their personal information with third parties \cite{jai2016privacy}. 


The relation between 
privacy concerns and gender has also been studied \cite{cho2009multinational}. \inlinecite{jai2016privacy} 
found that women were less willing than men 
to permit third parties to share their personal information. 
Similarly, \inlinecite{rowan2014observed} 
observed that women reported greater information privacy concerns than their male counterparts. Both studies considered gender as binary. 


Another relevant factor is participants' internet experience. 
As users grow in internet experience, concerns for online information privacy may decrease \cite{zukowski2007examining}. \inlinecite{bellman2004international} 
concluded that participants with more internet experience were less concerned about online privacy overall, and in particular, were less 
worried about \textit{improper access} and \textit{secondary use}. This could be explained 
by increased familiarity with online privacy practices 
\cite{zukowski2007examining}. 

\section{Research Questions}
\label{sec:research_question}

\hlreview{Overall, while concepts around information privacy concerns have been extensively investigated, some limitations are shared among the studies that assess differences in these concerns worldwide.} Most research has been conducted through surveys and has focused only on a few geographic regions, with a notable exception of \cite{li2017cross}. Many studies have had a limited sample size \cite{vitkauskaite2010overview,ur2013cross,ebert2020does,su12198286,doi:10.1080/0144929X.2020.1831608,OGHAZI2020531,krasnova2010privacy}\hlreview{; thus, their findings' generalizability} has been questioned \cite{lee2019information}. Moreover, when information privacy concerns questionnaires are delivered in English to speakers of other languages, key differences among countries may be obscured, as has happened with other cross-national research \cite{harzing2002interaction,harzing2006response}. Unfortunately, conducting larger-scale, multi-country, and multi-language surveys can be quite expensive \cite{harzing2005does,doi:10.1080/0144929X.2020.1831608}. Yet, large-scale research to deepen our understanding of information privacy concerns worldwide is still needed \cite{vitkauskaite2010overview,su12198286,doi:10.1080/0144929X.2020.1831608,va_zou2018ve,OGHAZI2020531,OKAZAKI2020458}.

\hlreview{We seek to assess the feasibility of using social media data to identify information privacy concerns and characterize language and regional differences.} Twitter is a popular micro-blogging service where individuals from different world regions who speak diverse languages share opinions, information, and experiences \cite{yaqub2017analysis,shen2015analysis}. Mining text from this platform has been used as a fast and inexpensive method to gather opinions from individuals \cite{o2010tweets}, which can complement findings obtained \hlreview{from traditional polls or other research methods}. Prior research has found a significant correlation between tweets and public opinion in diverse domains \cite{o2010tweets,tumasjan2010predicting,10.1145/3396956.3396973,doi:10.1080/21645515.2020.1714311}. 
\final{Following this trend of research, we aim to investigate whether Twitter data can reveal people's information privacy concerns. Thus, our first research question is as follows:} 
\begin{itemize}
    \item \hlreview{\textit{RQ1: Which information privacy concerns are present over 
    social media content about a data-breach scandal?}}
\end{itemize}

As we have reviewed in the prior section, there are arguments and evidence to support that information privacy concerns can vary across culture, language, and demographics
\cite{su12198286,doi:10.1080/0144929X.2020.1831608,OGHAZI2020531,OGHAZI2020531,gonzalez2019information}. \final{If information privacy concerns are present in a Twitter dataset, we could explore how they differ across people who live in different parts of the world and those who speak different languages. As we do not expect any specific trend of differences, we propose to test the following null hypotheses:}

\begin{itemize}
    \item \textit{ H0a. There are no differences in information privacy concerns by language}
    \item \textit{ H0b. There are no differences in information privacy concerns by world region.}

\end{itemize}


\section{Data \& Methods}
\label{sec:data_and_methods}

\begin{figure}[!h]
    \centering
    \includegraphics[width=\linewidth]{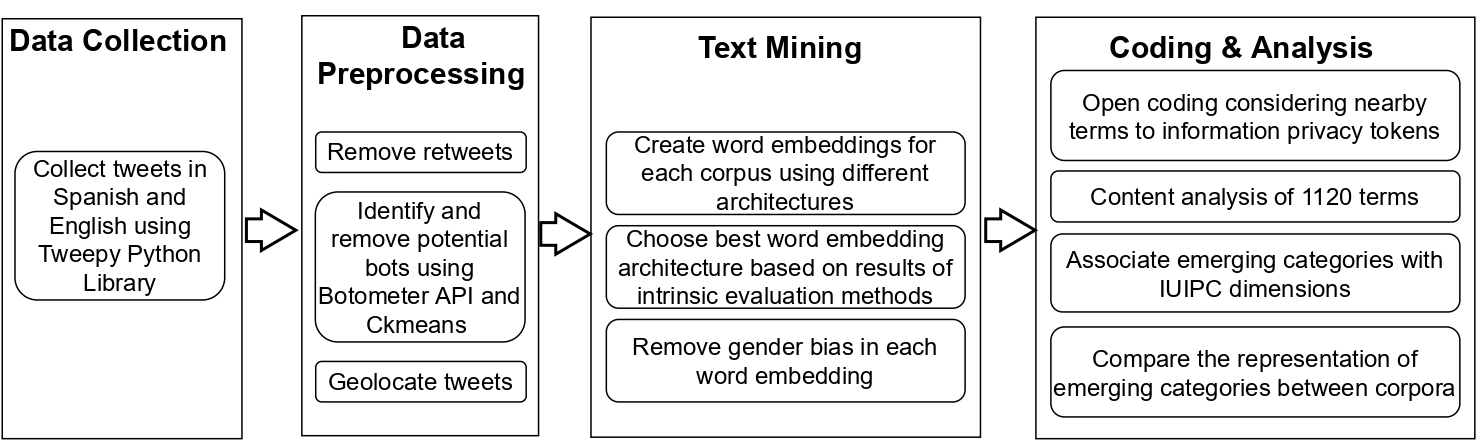}
\caption{Methodology flow chart}
\label{fig:methodology}
\end{figure}

To answer our research question \hlreview{and test the hypotheses,} 
we implemented a four-step methodology 
(see Fig \ref{fig:methodology}). We retrieved tweets associated with data privacy during a specific period (\textit{\ref{collection}. data collection}). We filtered the data, removing retweets and \hlreview{excluding} tweets likely to be generated by bots (\textit{\ref{preprocessing}. data pre-processing}). \final{We created word-embeddings (a multi-dimensional representation of a corpus) for the remaining tweets according to their language and world region} 
(\textit{\ref{mining}. text mining}). Finally, we conducted an analysis to identify similarities and differences in the semantic contexts of privacy keywords \hlreview{in the word embeddings }
(\textit{\ref{analysis}. coding and analysis}). Details about each of these steps are presented below.


\subsection{Data collection}
\label{collection}

We retrieved tweets related to the Facebook and Cambridge Analytica scandal between April 1st and July 10th, 2018. We focused on tweets in Spanish and English.

On March 17, 2018, it was revealed that the data firm Cambridge Analytica used personal data of 87 million Facebook users  for political advertising purposes without their consent \cite{schneble2018cambridge,OGHAZI2020531}. This scandal caused the closure of Cambridge Analytica \cite{solon2018cambridge} and numerous lawsuits against Facebook in the USA and the European Union. On Twitter, a \#DeleteFacebook campaign started as a response to this scandal \cite{lin2018deletefacebook}. \final{As the Cambridge Analytica scandal triggered Twitter users from different world regions (who speak diverse languages) to spontaneously share their opinions, experiences, and perspectives about data privacy, we decided to use a sample of these
tweets to answer our research question and test our hypotheses.}

We used Tweepy\footnote{http://www.tweepy.org/} to collect relevant tweets. Tweepy is a Python library for accessing the standard real-time streaming Twitter API,\footnote{https://developer.twitter.com/en/docs/tweets/filter-realtime/guides/basic-stream-parameters.html} which allows to freely retrieve tweets that match a given query. If the query is too broad that it includes over 1\% of the total number of tweets posted at that time worldwide, the query's response is sampled \cite{aghababaei2017activity,morstatter2014biased}. The way in which Twitter samples the data is unpublished. Nevertheless, studies have shown that as more data from the API is retrieved, a more representative sample of the
Twitter stream is obtained \cite{leetaru_is_nodate,morstatter2013sample}. 

\hlreview{To obtain relevant tweets, we used Tweepy's language filter to retrieve tweets in Spanish or English. We manually crafted a list of hashtags and keywords related to the Cambridge Analytica scandal. We collected tweets that had at least one of these terms. Examples of these terms are: ``\#DeleteFacebook'', ``\#CambridgeAnalytica'', ``\#Mark Zuckerberg'',``Facebook'', ``Facebook Cambridge'', 
and ``Facebook data breach''. Additionally, when appropriate, we added translations to Spanish of these terms to build the Spanish dataset.\footnote{
The authors are fairly confident of the quality of these translations because some of them are Spanish native speakers while others are English native speakers} 
In this way, if a tweet in Spanish had a hashtag in English, the tweet was collected and added to the Spanish dataset. A full list of the terms used to retrieve our data is available online\footnote{https://github.com/gonzalezf/Regional-Differences-on-Information-Privacy-Concerns}.}

Following this procedure, we retrieved more than $470,000$ tweets in Spanish and more than 7.4 million tweets written in English (see Table \ref{Table:NumberOfTweets}). The tweets in Spanish were produced by approximately 220,000 users while tweets in English were generated by about 1.8 million unique Twitter accounts.

\begin{table}[!h]
\caption{Datasets before and after data cleaning}
\label{Table:NumberOfTweets}
\begin{tabular}{@{\extracolsep{6pt}}lrrrr}
\toprule
Dataset                 & \multicolumn{2}{c}{Spanish}  & \multicolumn{2}{c}{English}              \\ 
    \cmidrule{2-3} 
    \cmidrule{4-5} 
                        & \#Tweets  & \#Accounts   & \#Tweets      & \#Accounts   \\
\midrule
Total                   & 472,363   & 222,352   & 7,476,988     & 1,846,542   \\
Original                & 106,656   & 47,951    & 1,572,371     & 574,452       \\
With Botometer score    & 100,606   & 44,182    & 1,442,112     & 504,214\\ 
Human-owned             & 74,644    & 36,056    & 975,678       & 410,180\\

\bottomrule
\end{tabular}
\end{table}


\subsection{Data pre-processing}
\label{preprocessing}
As we meant to analyze people's opinions about information privacy, we decided to pre-process our data in three ways. We removed all retweets to avoid analyzing exact duplicates. Afterwards, we sought to identify and filter out tweets that were generated by bots. Our last step was to associate tweets with different world regions. \hlreview{We further explain each of these steps below.}

\celeste{
First, we excluded retweets to avoid analyzing exact duplicates of content. This methodology step is suggested by several authors} \cite{HAJJEM2017761,aguero2021discovering,8963749}.
\celeste{
We kept tweets, quoted tweets, and replies to tweets. Exclusion of retweets reduced our datasets' size by 80\%. We refer to the resulting datasets as \textit{original} tweets (see Table {\ref{Table:NumberOfTweets}}).}

We used Botometer \cite{davis2016botornot} to detect and remove tweets created by bots.  
Botometer uses machine-learning 
to analyse more than one thousand features \cite{badawy2018analyzing} including tweets' content and sentiment, accounts' and friends' metadata, retweet/mention network structure, and time series of activity \cite{varol2017online,yang2019arming} to generate a score 
that ranges from 0 to 1. 
A higher value suggests a high likelihood that an inspected account is 
a bot \cite{badawy2018analyzing}. This tool  
has reached high accuracy (94\%) in predicting both simple and sophisticated bots
\cite{varol2017online,badawy2018analyzing}. Botometer is free and has been widely used\footnote{Since its release in May 2014, Botometer has served over one million requests \cite{davis2016botornot} via its website (https://botometer.iuni.iu.edu) and its Python API (https://github.com/IUNetSci/botometer-python)} 
 \cite{varol2017online,yang2019arming}.

\hlreview{Botometer processed all of the Twitter accounts who wrote original tweets. It}
returned a score for 44,182 (92.14\%) and 504,214 (87.77\%) 
\hlreview{accounts} of the Spanish and English datasets, respectively. \hlreview{Botometer cannot generate scores for suspended accounts or those that have their tweets protected. We decided to remove the tweets from these accounts from our datasets because we cannot confidently claim that they come from humans' accounts.} We applied the Ckmeans \cite{wang2011ckmeans} algorithm to define a threshold to distinguish between humans' and bots' accounts. 
For each language, we 
clustered the Botometer scores into five groups, where the first group included the accounts with the lowest scores (more human-like) and the fifth group \hlreview{comprised those} with the highest scores (more bot-like). After manually inspecting the accounts around the thresholds of each group, we concluded that the fourth and fifth groups in each dataset were unlikely to contain human accounts. Therefore, we used the fourth group's lowest threshold to discriminate humans' and bots' accounts. Accounts with a score lower than 0.4745 and 0.4947 in Spanish and English, respectively, were considered as human-owned. These thresholds are similar to those used in related work, where scores lower than 0.5 had been considered as humans \cite{varol2017online,badawy2018analyzing}. As a result, our datasets contain 36,056 human-owned accounts that created 74,644 tweets in Spanish and 410,180 accounts that created 975,678 tweets in English.

Finally, we used the GeoNames API\footnote{http://www.geonames.org/} to identify the country of residence of Twitter users in our datasets.  
On Twitter, users can self-report their city or country of precedence. Nevertheless, 
textual references to geographic locations can be ambiguous. For example, over 60 different places around the world are named ``Paris''
\cite{jackoway2011identification}. To deal with this challenge, we employed the GeoNames API, which is a collaborative gazetteer project that contains more than 11 million entries and alternate names for locations around the world in a variety of different languages \cite{bergsma2013broadly}. Given a text, its algorithm performs operations to recognize potential locations, followed by a disambiguation process. This last step checks hierarchical relations and picks 
a location by their proximity to other
locations mentioned in the text \cite{lambdaalphagammaovarsigma2017comparative}. 
This tool has yielded results with an accuracy above 80\% \cite{jackoway2011identification}.

We found that 80.68\% of users in our Spanish dataset and 78.68\% of users in our English dataset had filled the city or country fields in their profiles. 
However, the GeoNames API could not detect the users' location in several cases, for example when inaccurate information was provided (e.g., ``Planet Earth.. where everyone else is from'', ``Mars''). Nonetheless, the tool was able to identify the location of users who created 58.7\% of the Spanish tweets
and 59.9\% of the English ones. 
In the Spanish dataset, most tweets came from Spain (16.5\%) and Latin American countries, such as Mexico (11.9\%), Argentina (6.2\%), and Venezuela (4.7\%). In the English dataset, the majority of tweets came from the United States (32.4\%), followed by United Kingdom (6.9\%), India (3.2\%), and Canada (2.7\%) (see Table \ref{tab:geolocated_number_tweets}). 

\begin{table}[!h]
\caption{Top-10 
most frequent user locations in the Spanish and English datasets}
\label{tab:geolocated_number_tweets}
\begin{adjustbox}{max width=\textwidth}

\begin{tabular}{@{\extracolsep{3pt}}lrrrrlrrrr@{}}
\toprule
\multicolumn{5}{c}{Spanish} & \multicolumn{5}{c}{English} \\ 
\cmidrule{1-5}
\cmidrule{6-10}
& \multicolumn{2}{c}{Tweets}  & \multicolumn{2}{c}{Users} &  & \multicolumn{2}{c}{Tweets}  & \multicolumn{2}{c}{Users} \\
\cmidrule{2-3}
\cmidrule{4-5}
\cmidrule{7-8}
\cmidrule{9-10}

Location & \#  & \%  & \#  & \%  & Location & \#  & \% & \# & \%  \\
\midrule

Spain & 12,342 & 16.5 & 5,483 & 15.2 & U.S & 315,913 & 32.4 & 132,155 & 32.2 \\
Mexico & 8,852 & 11.9 & 4,720 & 13.1 & U.K & 66,901 & 6.9 & 29,656 & 7.2 \\
Argentina & 4,648 & 6.2 & 2,505 & 6.9 & India & 30,781 & 3.2 & 12,424 & 3.0 \\
Venezuela & 3,518 & 4.7 & 1,094 & 3.0 & Canada & 26,487 & 2.7 & 11,564 & 2.8 \\
Colombia & 2,447 & 3.3 & 1,348 & 3.7 & Australia & 13,375 & 1.4 & 6,501 & 1.6 \\
U.S & 1,823 & 2.4 & 948 & 2.6 & Germany & 9,260 & 0.9 & 3,493 & 0.9 \\
Chile & 1,806 & 2.4 & 1,073 & 3.0 & France & 8,605 & 0.9 & 3,006 & 0.7 \\
Peru & 1,116 & 1.5 & 619 & 1.7 & Nigeria & 5,156 & 0.5 & 2,787 & 0.7 \\
Ecuador & 893 & 1.2 & 455 & 1.3 & U.A.E & 5,120 & 0.5 & 1,504 & 0.4 \\
Brazil & 587 & 0.8 & 172 & 0.5 & South Africa & 4,962 & 0.5 & 2,912 & 0.7 \\
Other & 5,815 & 7.8 & 3,100 & 8.6 & Other & 98,100 & 10.1& 42,658 & 10.4 \\
Unknown & 30,797 & 41.3 & 14,574 & 40.4 & Unknown & 391,018 & 40.1 & 161,767 & 39.4 \\
\bottomrule
\end{tabular}
\end{adjustbox}
\end{table}

To compare information privacy concerns by geographical regions, we divided the Spanish Twitter dataset in two sets: tweets written by users from (1) Latin America and (2) Europe. Similarly, we categorized the English dataset into three groups: tweets written by users from (1) North America, (2) Europe and (3) Asia. As a result, five different language-regional datasets
were generated. The Spanish-Latin America dataset includes 27,839 tweets written by 13,937 users. The Spanish-Europe dataset comprises 12,799 tweets created by 5,774 accounts. Regarding the English data, the North America dataset includes 342,400 tweets generated by 142,719 users, the English-Europe one has 111,745 tweets 
of 46,927 users, and the English-Asia dataset contains 42,208 tweets produced by 17,929 accounts (Table \ref{tab:regional_number}). 
\celeste{We did not consider other subsets because of their small size. In Spanish, we only collected 1,929 tweets from North America 
and 217 tweets from Asia. 
In English, we only collected 3,851 tweets from Latin America. 
}

\begin{table}[!h]
\caption{Tweets and users in each dataset}
\label{tab:regional_number}
\begin{tabular}{@{}clrr@{}}
\toprule
\multicolumn{1}{l}{Language} & Region & \# of tweets & \# of users \\ \midrule
\multirow{2}{*}{Spanish} & Latin America & 27,839 & 13,937  \\
& Europe & 12,799 & 5,774 \\ 
\midrule
\multirow{3}{*}{English} & North America & 342,400 & 143,719  \\
& Europe & 111,745 & 46,927 \\
& Asia & 42,208 & 17,929 \\
 \bottomrule
\end{tabular}
\end{table}

\subsection{Text mining: Word embeddings to identify semantic contexts}
\label{mining}

We employed word embeddings \cite{mikolov2013distributed} to characterize the semantic context in which 
privacy-related keywords 
are framed. 
Based on co-occurrence of terms, word embeddings create a reduced multi-dimensional representation of a corpus. Such representation can be used to 
analyze 
the semantic proximity among the corpus' terms. 
Analyzing the closest terms of a given term can reveal the semantic context in which it is used \cite{rho2018fostering,gonzalez2019information}.

We created a set of word embeddings to enable cross-language and cross-regional comparisons. First, we built word embeddings for the Spanish and English datasets (containing 
both geolocated and non-geolocated tweets). Then, 
we generated word embeddings for each of 
our five language-regional datasets. 
Before creating the word embeddings, we transformed the text to lowercase. We also removed stop-words and digits from the tweets. We customized our stop-words to ensure that symbols like ``\#'' were removed but not the words that follow it. Links and usernames were removed. Words with total frequency lower than three were ignored. These steps downsized the vocabulary 
by approximately 67\%  (details in Table \ref{tab:vocabulary_size_datasets}). 

\begin{table}[!h]
\caption{Initial and final vocabulary size  in each  dataset}
\label{tab:vocabulary_size_datasets}
\begin{tabular}{@{}clrr@{}}
\toprule
\multicolumn{1}{l}{Language} & Region & Initial vocabulary size & Final vocabulary size\\ \midrule
\multirow{3}{*}{Spanish}  & All  & 65,036 &  21,736\\
\cmidrule{2-4} 
& Latin America  & 35,149 & 11,359 \\
 & Europe & 21,630 & 6,696 \\ 
 \midrule
 \multirow{4}{*}{English}  & All & 244,371 &  76,128\\
\cmidrule{2-4} 
& North America  & 115,710 & 41,109 \\
 & Europe & 66,042 & 23,514 \\
 & Asia  & 39,120 & 13,896 \\
 \bottomrule
\end{tabular}
\end{table}


We considered eight word embedding architecture combinations that involve \textit{Word2Vec/FastText}, \textit{CBOW/Skipgram} and different numbers of dimensions and epochs. As there is still no consensus 
about which word embedding evaluation method is more adequate 
\cite{bakarov2018survey}, we evaluated each word embedding architecture for the 
English dataset 
over 18 intrinsic conscious 
evaluation methods \cite{bakarov2018survey} 
using a word embedding benchmark library.\footnote{https://github.com/kudkudak/word-embeddings-benchmarks} 
\inlinecite{bakarov2018survey} approach  has categorized the 
evaluation methods 
in three categories: 
\begin{itemize}
  \item Word semantic similarity (WSS): RW, MEN, Mturk287, WS353R, WS353S, WS353, SimLex999, RG65 and TR9856
  \item Word Analogy (WA): Google Analogy Test set, MSR and SemEval 2012-2 
  \item Concept categorization (CC): AP, BLESS, BM, ESSLI 1A, ESSLI 2B, and ESSLI 2C
\end{itemize}

To choose the best architecture, we designed a point system 
to reflect the embeddings' performance. 
For each 
evaluation method, the word embedding with the highest accuracy received a score of 8 points, the embedding with the second highest accuracy was assigned 
7 points, 
and so on. 
After running all evaluation methods, we summed the points obtained for each architecture
. 
Considering a negative sampling and windows size parameters equal to 5, 
a Word2Vec CBOW architecture with 300 dimensions trained during 50 epochs achieved the total highest score 
(see Table \ref{tab:WordEmbeddingsEvaluation}). 
The same architecture had the best performance for all English regional datasets. 
Given that these 
evaluation methods are not available for a Spanish corpus, the same architecture was used to create 
all the Spanish word embeddings.

\begin{table}[!h]
\caption{Word embedding architectures and their evaluation scores.  Best performance is indicated with bold font style.}
\label{tab:WordEmbeddingsEvaluation}

\begin{tabular}{@{\extracolsep{6pt}}llrrrrrr@{}}
\toprule
\multicolumn{4}{c}{Architecture} & \multicolumn{4}{c}{Evaluation word embedding score} \\ 
\cmidrule{1-4} 
\cmidrule{5-8} 
Type & Model & Dim. & Epochs & WSS & WA & CC & Total \\\midrule
FastText & CBOW & 100 & 10 & 22 & 19 & 25 & 66 \\
Word2Vec & Skipgram & 100 & 10 & 40 & 4 & 35 & 79 \\
Word2Vec & CBOW & 100 & 10 & 35 & 11 & 33 & 79 \\
Word2Vec & CBOW & 100 & 50 & 34 & 16 & 41 & 91 \\
Word2Vec & CBOW & 100 & 300 & 33 & 9 & 31 & 73 \\
Word2Vec & CBOW & 300 & 10 & 40 & 18 & 32 & 90 \\
\textbf{Word2Vec} & \textbf{CBOW} & \textbf{300} & \textbf{50} & \textbf{53}& \textbf{21} & \textbf{36} & \textbf{110} \\
Word2Vec & CBOW & 300 & 300 & 31 & 10 & 29 & 70 \\ \bottomrule
\end{tabular}

\end{table}

Previous work has reported that word embeddings can reflect gender bias as a result of social constructs embedded in the data \cite{zhao2018learning,jha2017does}. 
To reduce gender bias while preserving its useful properties such as the ability to cluster related concepts, we followed \inlinecite{bolukbasi2016man} approach. This is a post-processing method that projects gender-neutral words 
to a subspace which is perpendicular to a 
gender dimension, defined by a set of 
terms associated with 
gender 
such as \textit{girl}, \textit{boy}, \textit{mother} and \textit{father} 
\cite{zhao2018learning}. We applied the following procedure to our English embeddings: (1) we identified a gender subspace selecting pairs of English words that can reflect a gender direction in each word embedding such as \textit{woman-man}, \textit{daughter-son} and \textit{female-male}, (2) we ensured that gender neutral words are zero in the gender subspace, and (3) we made neutral words equidistant to all pair of terms contained in a collection of equality sets. A equality set is composed by a pair of words that should differ only in the gender component such as \textit{\{grandmother, grandfather\}} and \textit{\{guy, gal\}}. During this 
process, we used 
the English terms suggested by \inlinecite{bolukbasi2016man}. 
For the Spanish word embeddings, we used Google Translate API\footnote{https://cloud.google.com/translate/} to translate the same terms. 

\subsection{Manual coding \& analysis}
\label{analysis}


We \hlreview{conducted a systematic qualitative examination of the semantic contexts in which information privacy terms appear according to the word embeddings. First, we conducted open coding 
of the semantic neighborhoods of privacy-related keywords. After several iterations, we developed a set of categories to characterize them. To assess if information privacy concerns were present (RQ1), we contrasted these categories to a widely accepted framework to describe internet users' information privacy concerns.} 
%

We focused our investigation on four keywords in English: \textit{information}, \textit{privacy}, \textit{users} and \textit{company}. We used their corresponding translations in Spanish: \textit{informaci\'{o}n}, \textit{privacidad}, \textit{usuarios} and \textit{empresa}. \hlreview{
We chose to include \textit{information} and \textit{privacy} because they are the main concepts under study. We could have added data; however, its semantic context is almost identical to that of information. Thus, adding it would have resulted in a mere duplication of terms. To increase the size of our dataset, we decided to add 
\textit{users} and \textit{company} because of their key roles in respect of controlling and safeguarding personal information. We also considered these terms more specific to the vocabulary of the data privacy domain than alternative ones (e.g., people, organizations)}. 

For each embedding, we retrieved the closest terms to the four keywords. 
Closeness between each term 
and 
a keyword was measured using cosine similarity. For instance, 
the closest terms retrieved to the keyword \textit{information} in the English word embedding were \textit{info}, \textit{data}, \textit{details}, and \textit{personal}, in that order. \hlreview{We chose to study the 40 closest terms after careful examination of the lists of close terms according to our different embeddings. After the 40th position in these lists, we rarely found terms that were even slightly related to information privacy. We reason that the value of this threshold is dataset-dependent. It is likely to be related to the vocabulary sizes (ours range from 6,696 to 41,109). In our case, we opted for using 40 as the threshold to study the semantic context of each keyword. Hence, we qualitatively analyzed 160 terms for each embedding. Overall, our dataset for qualitative analysis included 1,120 terms.} 

Two of the authors conducted open coding of the 320 terms retrieved from the Spanish and English word embeddings. Open coding is a process to identify, define and develop categories based on properties and dimensions of raw data \cite{williams2019art}. We used this technique to identify distinct concepts and themes 
from the extracted terms \cite{williams2019art}. After inspecting the retrieved terms during several iterations, the coders developed a coding guideline with multiple concept categories and their corresponding explanations to classify the retrieved terms. For example, the term \textit{info} extracted from the keyword \textit{information} was categorized as a \textit{synonymous}, given that we can attribute to it the same meaning. 
The terms \textit{data} and \textit{details} were classified as \textit{data} \& \textit{information}, and \textit{personal} was labeled as \textit{attribute or characteristic}. During a series of meetings, both coders compared their categorization process and refined a common coding guideline, establishing rules that would increase the categorization's reliability. The goal during this process is to segregate, group, regroup and re-link the terms to consolidate meaning and explanation of the categories \cite{williams2019art}. 

At the end of this process, 15 categories emerged from the data (see Table \ref{tab:coding_guideline}). Considering the four keywords, an inter-coder reliability measure (Cohen's kappa) of 0.685 and 0.754 were obtained for the Spanish and English dataset, respectively. These scores indicate substantial agreement \cite{viera2005understanding} during the process.

We repeated the procedure 
for the regional datasets. 
The coders categorized the 40 closest terms to the keywords 
according to the coding guideline. 
Through an iterative process, a total of 800 words were manually coded. 
No new categories emerged from the data. 
On average, a Cohen's kappa above 0.722 
was obtained in all the regional datasets.






\begin{table}[!h]
\caption{Inter-rater reliability (Cohen's kappa) score by dataset }
\label{tab:kappa_score}
\resizebox{\textwidth}{!}{
\begin{tabular}{@{}clrrrrr@{}}
\toprule

\multicolumn{1}{c}{Language} & Region & Information & Privacy & Company & Users & \textbf{Avg. by dataset}\\ 

\midrule
\multirow{3}{*}{Spanish}  & All  & 0.864 & 0.630 & 0.604 & 0.642 & \textbf{0.685}\\
& Latin America & 0.749 & 0.673 & 0.687 & 0.778 & \textbf{0.722} \\
 & Europe  & 0.820 & 0.710 & 0.774 & 0.827 & \textbf{0.783}\\
 \midrule
 \multirow{4}{*}{English}  & All & 0.768 & 0.747 & 0.672 & 0.829 & \textbf{0.754}\\
& North America   & 0.831 & 0.721 & 0.912 & 0.971 & \textbf{0.859} \\
 & Europe & 0.777 & 0.743 & 0.805 & 0.807 & \textbf{0.783} \\
 & Asia  & 0.832 & 0.685 & 0.736 & 0.833 & \textbf{0.771} \\
 \midrule
 \multicolumn{2}{c}{\textbf{Average of all embeddings}}  &\textbf{0.806}&\textbf{0.701}&\textbf{0.741}&\textbf{0.812}&\textbf{0.765} \\
 \bottomrule
\end{tabular}
}
\end{table}

\hlreview{To assess if information privacy concerns were present in a Twitter dataset about a data-breach scandal (RQ1), we compared the resulting categories with the IUIPC's dimensions: \textit{collection}, \textit{control}, and \textit{awareness}. IUIPC is a theory-based model that has been widely used to study information privacy concerns on the internet} (see Section \ref{iuipc}). \celeste{Then, we tested the null hypotheses about differences in information privacy concerns across language and world regions (H0a and H0b). To do so, we used a Chi-squared test to assess if the proportion of terms in the semantic contexts were significantly different across word embeddings.
In all of these tests, we accounted for multiple comparisons by applying alpha adjustment according to Šidák } \cite{doi:10.1080/01621459.1967.10482935,article_sidak}.\celeste{ This method allowed us to control the probability of making false discoveries when performing multiple hypotheses tests.} 


\section{Results}
\label{sec:results}
\hlreview{
In this section, we address
our research question 
and 
test the null hypotheses about differences in information privacy concerns by language and world regions.}

As explained \hlreview{above}
, we create word embeddings for our Spanish and English datasets of tweets about the Cambridge Analytica scandal. Then, we take a closer examination of how the semantic context of four keywords varies across language and world regions. The semantic context is operationalized as the 40 closest terms of each keyword: \textit{information}, \textit{privacy}, \textit{company}, and \textit{users}. As an example, Table \ref{tab:most_similar_terms_company_users} and Table \ref{tab:most_similar_terms_information_privacy} 
show the 20 closest terms to the keywords, according to the Spanish and English word embeddings.\footnote{\celeste{Terms in Spanish were translated to English by the authors} 
}
Full results are available online.\footnote{https://github.com/gonzalezf/Regional-Differences-on-Information-Privacy-Concerns}

\begin{table}[!h]
\caption{Top 20 closest terms to \textit{information} and \textit{privacy} in the Spanish and English word embeddings}
\label{tab:most_similar_terms_company_users}
\begin{tabular}{@{\extracolsep{6pt}}llll}
\toprule
\multicolumn{2}{c}{Information} &
\multicolumn{2}{c}{Privacy} \\
\cmidrule{1-2}
\cmidrule{3-4}
Spanish & English & Spanish & English \\
\midrule
data & info & intimacy & data privacy \\
info & data & data & gdpr \\
third parties & details & confidentiality & protection \\
fact & personal & scams & users \\
third & users & personal data & user \\
interviewer & profiles & privacy policy & consumers \\
facebook & identifiers & digital security & data \\
users & personal data & data protection & transparency \\
privacy & private & identity & facebook \\
consent & records & minor & personal \\
authorization & user & facebook & consent \\
purposes & consent & third parties & security \\
private & advertisers & information & sharing \\
personal & permission & cibersecurity & data protection \\
location & metadata & sensitive & tos \\
ecomlancer & datas & emails & consumer \\
serve & companies & cookies & collection \\
viatec & individuals & protect yourself & opt \\
intimate & freely & suppose & trust \\
profiles & informations & take care of your data & privacyrights \\

\bottomrule
\end{tabular}
\end{table}
    
\begin{table}[!h]
\caption{Top 20 closest terms to `\textit{company} and \textit{users} in the Spanish and English word embeddings}
\label{tab:most_similar_terms_information_privacy}
\begin{tabular}{@{\extracolsep{6pt}}llll}
\toprule
\multicolumn{2}{c}{Company} &
\multicolumn{2}{c}{Users} \\
\cmidrule{1-2}
\cmidrule{3-4}
Spanish & English & Spanish & English \\
\midrule
company & companies & third parties & user \\
consultant & firm & sensitive & consumers \\
firm & companys & citizens & personal \\
organization & platform & illegally & peoples \\
obtain & firms & users & subscribers \\
relation & organization & authorization & customers \\
researcher & data & profiles & people \\
deliver & entity & private & facebook \\
plot & giant & used & data \\
finance & user & people & fb \\
ltd & facebook & illegal & apps \\
way & corporation & clients & individuals \\
facebook & fb & user & advertisers \\
illegally & organisation & improperly & privacy \\
companies & business & obtained & information \\
ca & users & nametests & app \\
own & businesses & information & citizens \\
brand & service & facebook & profiles \\
decide & ca & data & companies \\
scl & site & voters & collected \\
creole & personal & cambridgeanalytics & private \\
relations & organizations & infringement & consent \\
cambridge & employees & use & accounts \\
data & agency & purposes & permissions \\
laboratories & co & authorized & use\\
\bottomrule
\end{tabular}
\end{table}

\subsection{\hlreview{Information privacy concerns present in a Twitter dataset}}

\begin{figure}[!h]
    \centering
    \includegraphics[width=0.93\linewidth]{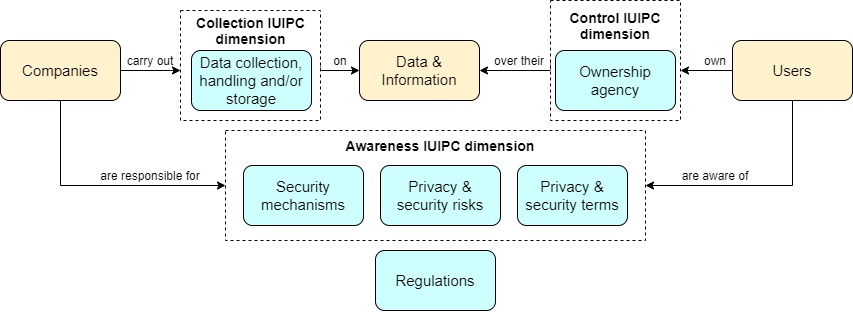}
    \caption{Relationships between our categories and IUIPC dimensions}
    \label{fig:iuipc_dimensions_per_language_details}
\end{figure}

As a result of the coding process, we define 15 categories to \hlreview{analyze the closest terms} 
(see Table \ref{tab:coding_guideline}). To answer our first research question 
,
we compare our categories with IUIPC, a framework widely used to measure information privacy concerns in the context of the Internet \cite{liu2018impact}. We find relationships among some of our categories and the three IUIPC concepts as well as our initial keywords, as shown in Figure \ref{fig:iuipc_dimensions_per_language_details}.

\begin{table}[!ht]
\setlength{\tabcolsep}{2pt} 
\renewcommand{\arraystretch}{1}
\scriptsize
\centering
\caption{\hlreview{Coding guideline to classify the semantic contexts of our keywords}. \celeste{Yellow background is for categories that match our initial keywords. Light blue background is for categories related to IUIPC. Gray background is for other categories.}} 
\label{tab:coding_guideline}
\begin{tabular}{p{0.16\textwidth}p{0.26\textwidth}p{0.27\textwidth}p{0.27\textwidth}}
\toprule
Categories & Description &Spanish Examples & English Examples\\ \bottomrule
\rowcolor{Yellow}
Data \& Information & 
Direct references to these concepts and examples of user data and its meaning
&
records, location, data, emails, profiles, contacts
& profile, messages, documents, location, accounts, metadata
\\
\rowcolor{Yellow}
Companies & 
Entities that manipulate user data for their own purposes
&facebook, cambridgeanalytica, scl, grindr, advertisers
& facebook, cambridgeanalytica, google, emerdata, apple
\\
\rowcolor{Yellow}
Users & Data owners &
users, consumers, population, citizens, people
& users, consumers, subscribers, citizens, people
\\
\rowcolor{LightCyan}
Data collection, handling and/or storage & Mechanisms and verbs associated with obtaining, collecting and handling data&
use, obtained, log in, collecting, apps, mechanisms
&collected, store, access, databases, usage, analyzed\\
\rowcolor{LightCyan}
Ownership agency & 
User-control over personal information&
authorization, agree, consent, protect
& consent, opt,  shield, autonomy
\\
\rowcolor{LightCyan}
Privacy \& security terms & 
Words associated with data privacy and security
&
cybersecurity, confidentiality, intimacy, safe, secure
& confidenciality,  transparency, privately, dataprivacy, security
\\
\rowcolor{LightCyan}
Security mechanisms & 
Tools and techniques that implement security services&
credentials, password, biometry, key

& encrypted, password,  biometric, privacybydesign
\\
\rowcolor{LightCyan}
Privacy \& security risks & 
Entities or bad practices that can compromise sensitive data&
trojan, cybercriminal, illegally, scams, stealing
& grooming, databreach, misuse, illegally, violated
\\
\rowcolor{LightCyan}
Regulation & Law, rule or regulation that controls the use of user data&
rgpd, right to be forgotten, privacypolicy, iso, habeasdata
& gdpr,  tos,  hippa, privacyrights, regulation
\\
\rowcolor{Gray}
Synonymous & Same meaning than the keyword&
info, company, firm, user, private
& info, companies, firm, privacy, user, 
\\
\rowcolor{Gray}
Attribute or characteristic&A characteristic of the keyword & 
false, private, external, specialized, britain
& sensitive, giant, holistic, strategic, affected
\\
\rowcolor{Gray}
Action & Action or activity linked to the keyword&

define, explode, promote, attend, move
& order, solve,  reveal, update, confirming
\\
\rowcolor{Gray}
Third party & 
Can not be categorized as User or Company. There is not sufficient contextual information to do so&
rrhh, medicians, interviewer, ex employee, philippine
& government, agency, indians, europeans, developers, \\
\rowcolor{Gray}
Reaction or attitude &
Way of feeling or acting toward a entity 
&
guilt, honest, overfall, suffers, handle
&
 willingly, freely, tighter, restricting, forced
\\
\rowcolor{Gray}
Undetermined & Relationship between keyword and term is unknown 
& 
approximately, v.i.p, ground, higher, depth
& psychological, millions, new, group, image 
\\
\bottomrule
\end{tabular}

\end{table}

Three categories match our initial keywords (Table \ref{tab:coding_guideline}, yellow background): (1) \textbf{data \& information} is associated with the \textit{information} keyword, including direct references to this concept and examples of user data and its meaning (e.g.,``messages'' and ``metadata''), (2) \textbf{companies} 
include terms about organizations that use personal data for their own purposes such as ``Facebook'' and ``Apple'', and (3) \textbf{users} contain references to this keyword (e.g., ``customers'', ``people'').

Five categories are related to IUIPC (Table \ref{tab:coding_guideline}, light blue background). We identify a \textbf{data collection, handling and/or storage} category that contains words associated with technology or techniques useful to obtain, collect or handle data (e.g: ``databases'', ``services'', ``app'', ``website''). This matches to the IUIPC’s \textit{collection} dimension, which refers to the ``degree to which a person is concerned about the amount of individual-specific data possessed by others relative to the value of benefits received''\cite{malhotra2004internet}. \celeste{Examples of tweets that include terms that fit this category are: 
}
 
 
 \begin{quotation}
 `\textsf{$@$hidden\_username $@$hidden\_username This is bigger than facebook because all social media outlets collect and store this data on every user. If you haven't looked, check and see what twitter has collected on you. Free apps are not free, neither are paid ones}'
 \end{quotation}
 \begin{quotation}
  `\textsf{Facebook collects and sells PII data. Google and others maintain behavioral data anonymously and serve ads against it, but don't connect that data to identities that are sold to advertisers. I was not aware Facebook was such an anomaly.}'
 \end{quotation}
 
 

The IUIPC's \textit{control} dimension denotes concerns about control over personal information. This is often exercised through approval, modification and opportunity to opt-in or opt-out \cite{malhotra2004internet}. Terms related to this dimension appeared in the coding phase (e.g., ``consent'', ``opt'', ``permission'') and were categorized as \textbf{ownership agency}. This category also includes advice directed to users and good privacy practices terms (e.g., ``prevent'', ``protect''). \hlreview{Examples of tweets related to this category are:}
 
 
 \begin{quotation}
 `\textsf{If anything we should learn from the \#Facebook data breach. Don't volunteer information and prevent that secondary data collection by using \#adblocker  and \#VPN}'
 \end{quotation}

 \begin{quotation}
  `\textsf{Cambridge Analytica whistleblower Christopher Wylie urges U.S. senators to focus less on data consent and more on the idea that it's almost impossible to opt out of, for example, Google.}'
 \end{quotation}
 
 
The third IUIPC's dimension is \textit{awareness}, which refers to individual concerns about her/his awareness of organizational information privacy practices \cite{malhotra2004internet}. Three of our categories are associated with this dimension: (1) \textbf{privacy and security terms} that include words associated with data privacy and security such as ``confidentiality'', ``transparency'' and ``safety''; (2) \textbf{security mechanisms} that refers to tools and techniques that implement security services (e.g., ``password'', ``encryption''); and, (3) \textbf{privacy \& security risks} that denote entities or bad practices that can compromise sensitive data, for example: \textit{``troyano''} (trojan), ``databreach'', ``grooming'' and \textit{``ciberdelincuente''} (cybercriminal). \hlreview{Tweets that use these terms are:} 

 \begin{quotation}
`\textsf{Hmm- what do you think? I forsee a wave of new social network startups- will any be able to rise? Besides privacy and transparency what else would you want from a social network? \#swtech}'
 \end{quotation}
 
  \begin{quotation}
`\textsf{WhatsApp Co-Founder To Leave Company Amid Disagreements With Facebook. Facebook's desire to weaken WhatsApp's encryption and collect more personal data reportedly fueled the decision}'
 \end{quotation}
 
 
 \begin{quotation}
 `\textsf{Canadian federal privacy officials warned that third-party developers' access to Facebook users' personal information raises serious privacy risks back in 2009. $@$hidden\_link}' \end{quotation}

 

Another privacy-related category emerges from our coding but can not be \hlreview{easily} associated with an IUIPC dimension. This is the \textbf{regulation} category, which includes terms associated with laws and rules that control the use of personal data such as ``gdpr'' in reference to the European General Data Protection Regulation or ``tos'' in reference to Terms of Services. \hlreview{Examples of tweets with these terms are:}

  \begin{quotation}
`\textsf{New regulation in Europe called gdpr makes companies liable for data breaches with penalties which include fines of a percentage of global turnover. It feels like all Zuckerberg is liable for is a slap on the wrist and having to apologise in public}'
 \end{quotation}
 
   \begin{quotation}
 `\textsf{\#Today we are confirming that multiple snippets of data from CI that was lifted from facebook are in Russia. If you are an EU citizen this means you have a right to sue both companies for gdpr based infringements.  We will be leading this cause should no one else step up....}'
 \end{quotation}
 
   \begin{quotation}
`\textsf{Senator to \#Zuckerberg: Your terms of services are only a few pages long. People complain when online contracts are too long and filled with legalese. Now lawmakers are complaining they're too short. What's the threshold for length and detail, and how do we decide?}'
 \end{quotation}
 

Other categories are identified as well (Table \ref{tab:coding_guideline}, gray background). The \textbf{attribute or characteristic} category contains 
modifiers of a specific keyword. For example, the term ``sensitive'' emerges from the closest terms to 
\textit{information},
and the term \textit{``britanica''} (British) appears among 
the nearest terms to 
\textit{company}.
The \textbf{action} category includes words related to an act. 
For instance, the verbs \textit{``obtener''} (obtain) and \textit{``entregar''} (deliver) come out among the closest terms to \textit{company}.
The \textbf{third party} category contains terms related to entities that can not be categorized as user or company because there is not sufficient contextual information to do so, such as ``indians'', ``third'', ``individuals'' and ``americans''. Additionally, the \textbf{reaction or attitude} category comprises terms that represent a way of feeling or acting toward a person, thing or situation. For example, the terms ``deny'' and ``admitted'' are present in the closest terms to
\textit{company}. 
A \textbf{synonymous} category emerge during the process as well. This contains equivalent terms to each keyword. For example, the terms ``info'' and ``informations'' are close to \textit{information} and the terms ``companys'', ``corporation'', ``companies'' and ``firm'' appear among 
the closest terms to 
\textit{company}.
Terms with no clear relation to the keywords were classified as \textbf{undetermined}.

\begin{figure}[!h]
    \centering
      \includegraphics[width=0.85\linewidth]{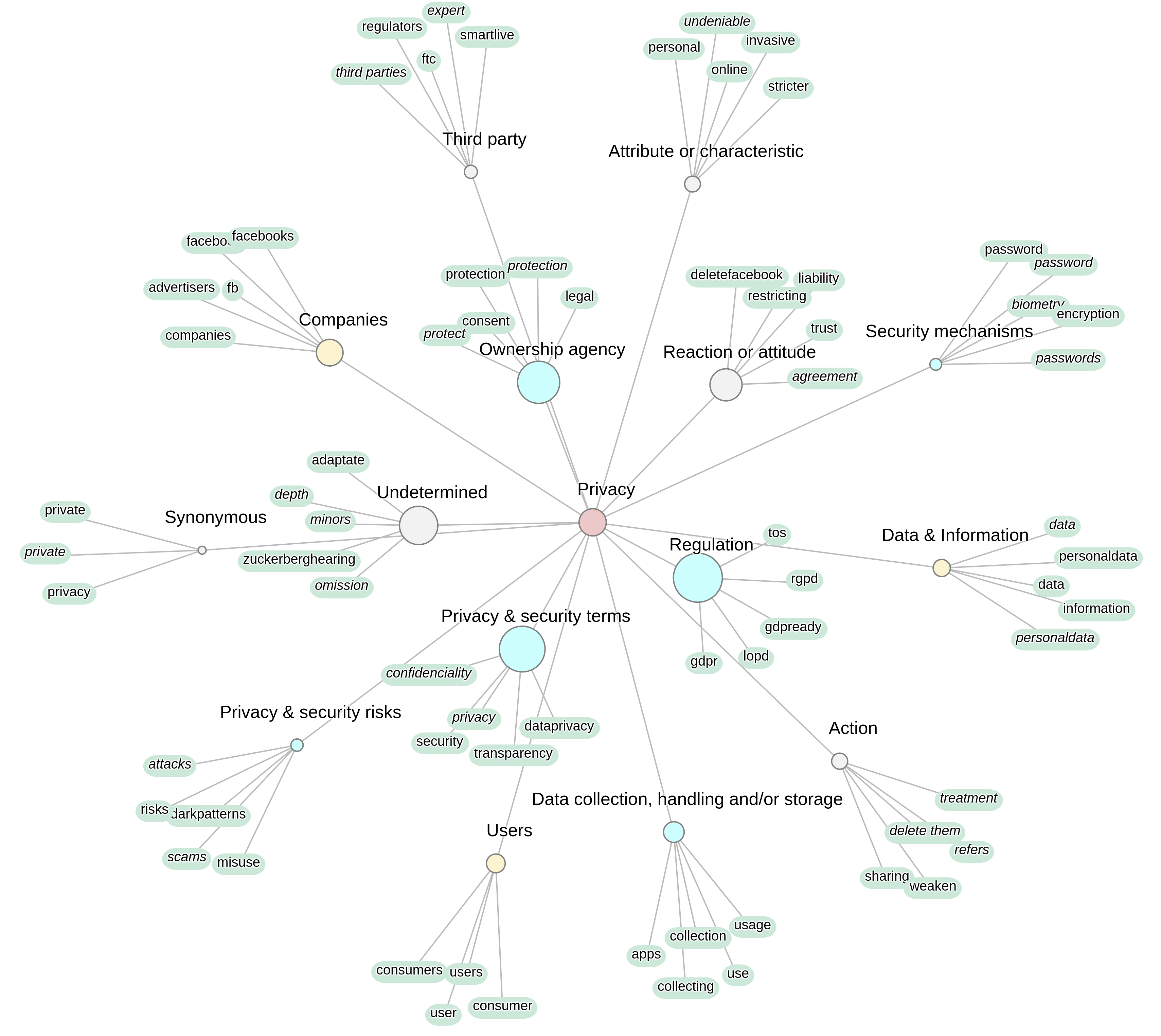}
\caption{This force-directed graph represents the open coding categories related to the keyword \textit{privacy} and provides examples of the terms that were coded as each category. Categories with the higher frequency are larger and closer to the keyword. Yellow nodes represent keywords and light-blue nodes denote privacy-related categories.}
\label{fig:closest_terms_to_privacy}
\end{figure}

We used force-directed graphs \cite{kobourov2004force} to represent all the categories that emerged from the analysis of semantic context of each keyword. Figure \ref{fig:closest_terms_to_privacy} shows the categories related to the keyword \textit{privacy}. In this graph, distance represents closeness in the semantic context. For example, terms that were categorized as regulation are closer to privacy than terms that were categorized as security mechanisms. Additionally, the visualization shows examples of terms in each category in Spanish or English.\footnote{Terms in Spanish were translated to English by the authors. These terms are shown in cursive} Force-directed graphs of the categories and terms associated with the other keywords (\textit{information}, \textit{company}, and \textit{users}) are available online.\footnote{\url{https://andreafigue.github.io/word_embeddings/visualization.html}}

\hlreview{Overall, we observe that the semantic contexts of four privacy-related keywords include terms corresponding to information privacy concerns. We illustrate such presence in Figure} \ref{fig:iuipc_dimensions_per_language_details}. \hlreview{We positioned each IUIPC dimension at the intersection between two of our keywords.} \textit{Companies} carry out collection, handling and or storage activities regarding \textit{data} \& \textit{information}. \textit{Users} exercise (some) agency over the control of their \textit{data} \& \textit{information}. The awareness dimension arises from the \textit{users}' perception of the \textit{companies}' practices. \hlreview{Our results suggest that the awareness dimension might be further categorized into sub-topics, such as awareness of} privacy and security terms, security mechanisms, and privacy and security risks.

Beyond what the IUIPC model proposes, we find that \textit{regulations} are relevant to Twitter users who talk about Cambridge Analytica. We position this concept close to \textit{awareness}, as it is considered an environmental factor that relates to information privacy concerns \cite{lee2019information,va_zou2018ve,MOHAMMED2017254} \hlreview{but is not integrated into the IUIPC.}

\subsection{Emphasis on information privacy concerns across languages and world regions}

\begin{figure}[!h]%
    \centering
    \subfloat[\centering By Language]{{\includegraphics[width=0.49\linewidth]{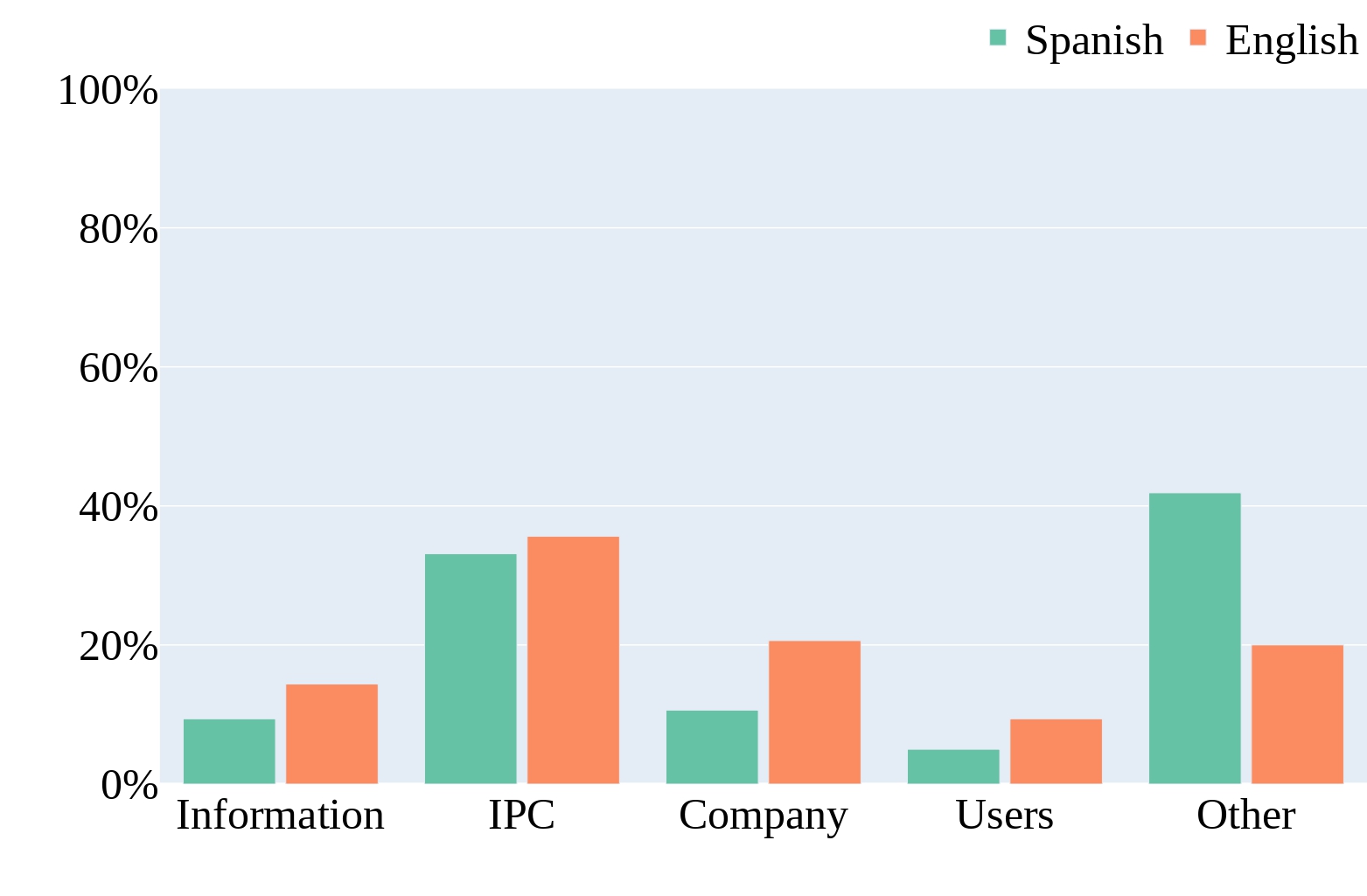} }}%
    \hfill
    \subfloat[\centering By world-region]{{\includegraphics[width=0.49\linewidth]{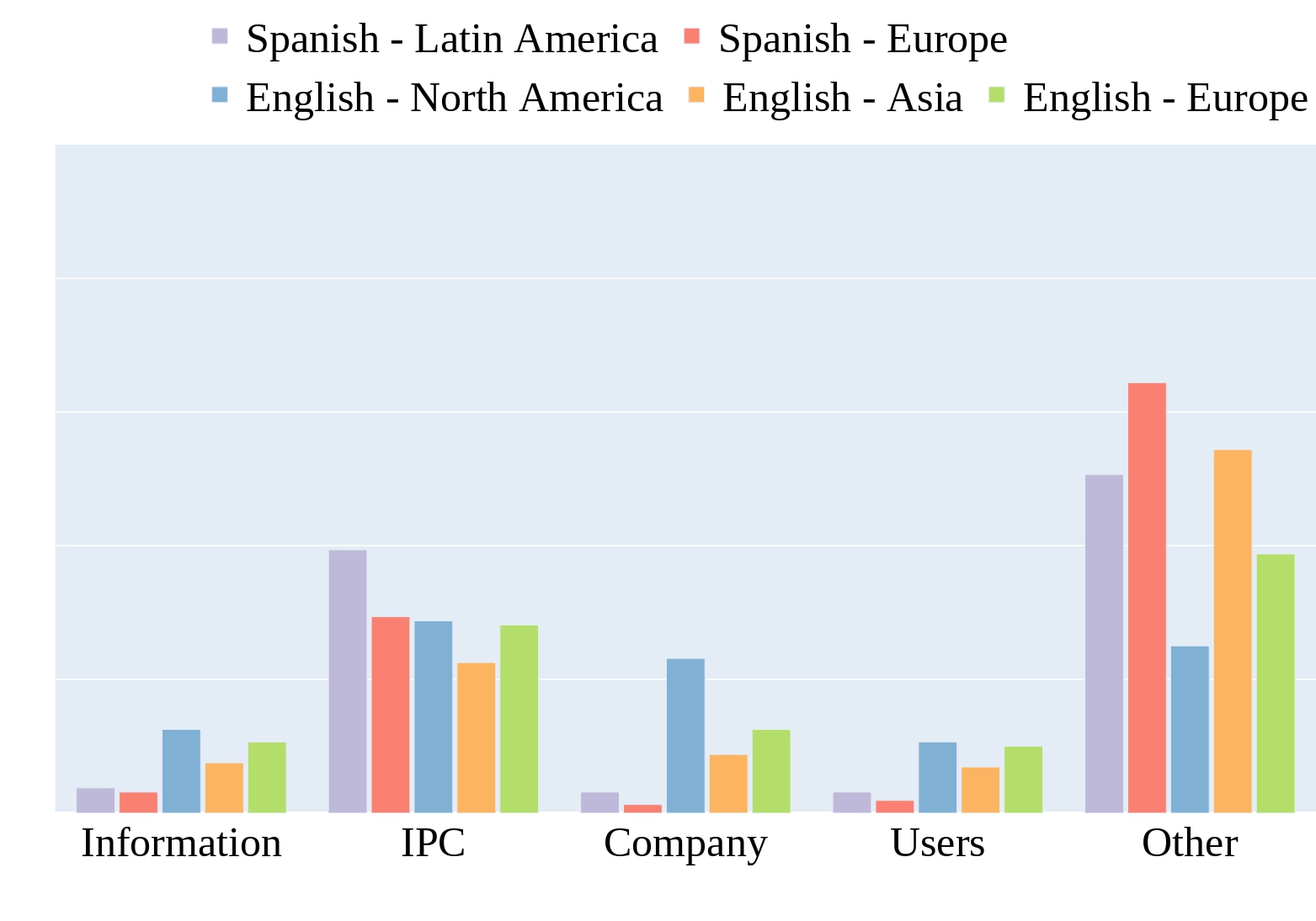} }}%
    \caption{Proportion of categories by language (a) and world region (b)}%
    \label{fig:global_categories_by_language_region}
\end{figure}

\hlreview{As we are able to observe the presence of information privacy concerns on the Twitter datasets, we can now turn to test the null hypotheses regarding differences across language and world regions.}
We compare the \hlreview{emphasis on} 
\hlreview{information privacy concerns (IPC)} in the semantic contexts that emerge from the different word embeddings. Figure \ref{fig:global_categories_by_language_region} reports the distribution of terms that relate to the initial keywords (Table \ref{tab:coding_guideline}, yellow background) and IPC (Table \ref{tab:coding_guideline}, light blue background) in each language and world region under study. \textit{Others} include all remaining categories. 

\celeste{To test our hypotheses, we performed Chi-square goodness-of-fit tests. Because we ran multiple tests, we applied Šidák  correction to counteract the problem of multiple comparisons, thus controlling the family-wise error rate. According to our Šidák ’s adjustment, to maintain an overall alpha of 0.05 for the collection of 10 tests, null hypotheses can be rejected when  $p < 0.0102$. }


\begin{figure}[!h]%
    \centering
    \subfloat[\centering By Language]{{\includegraphics[width=0.49\linewidth]{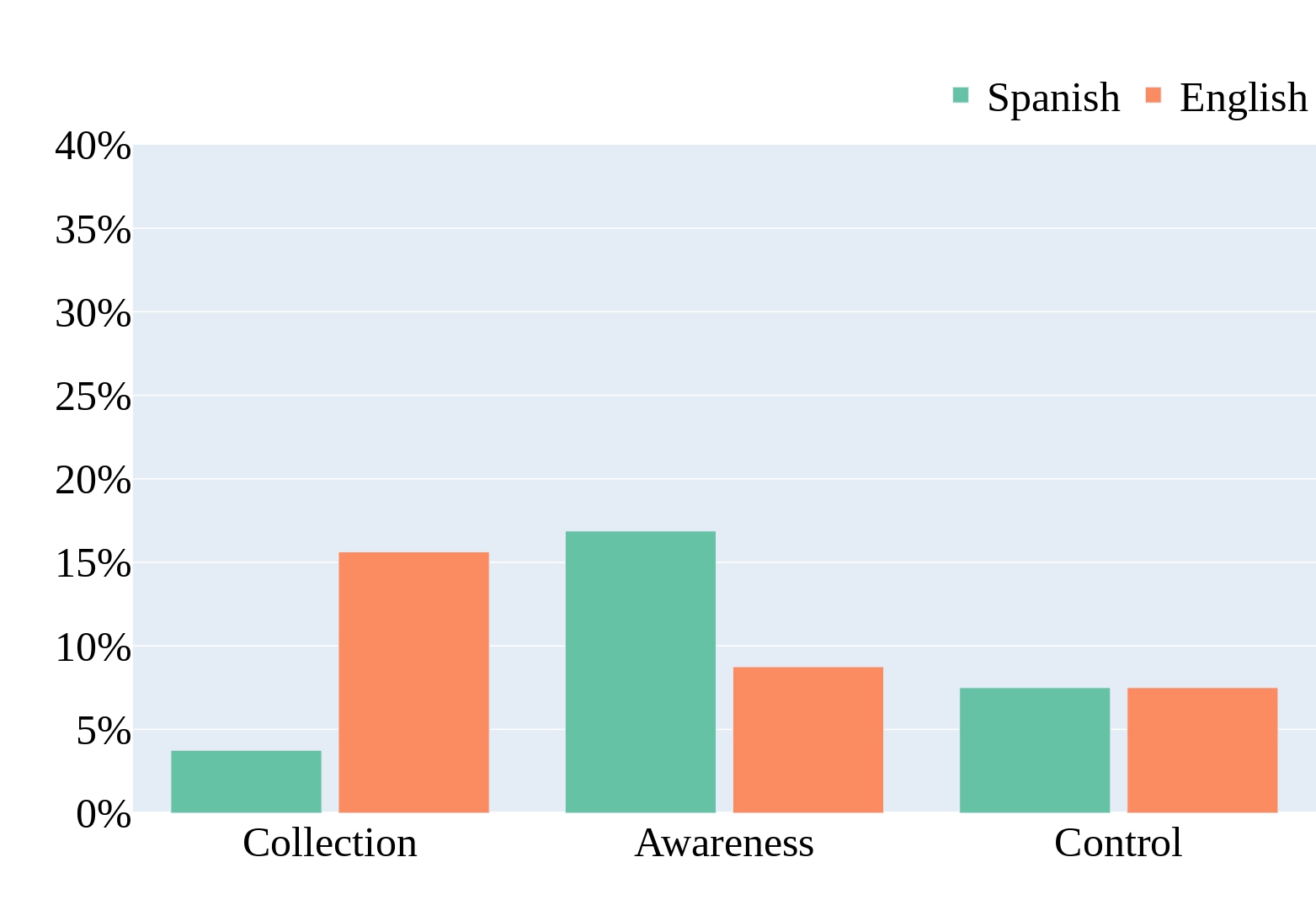} }}%
    \hfill
    \subfloat[\centering By world-region]{{\includegraphics[width=0.49\linewidth]{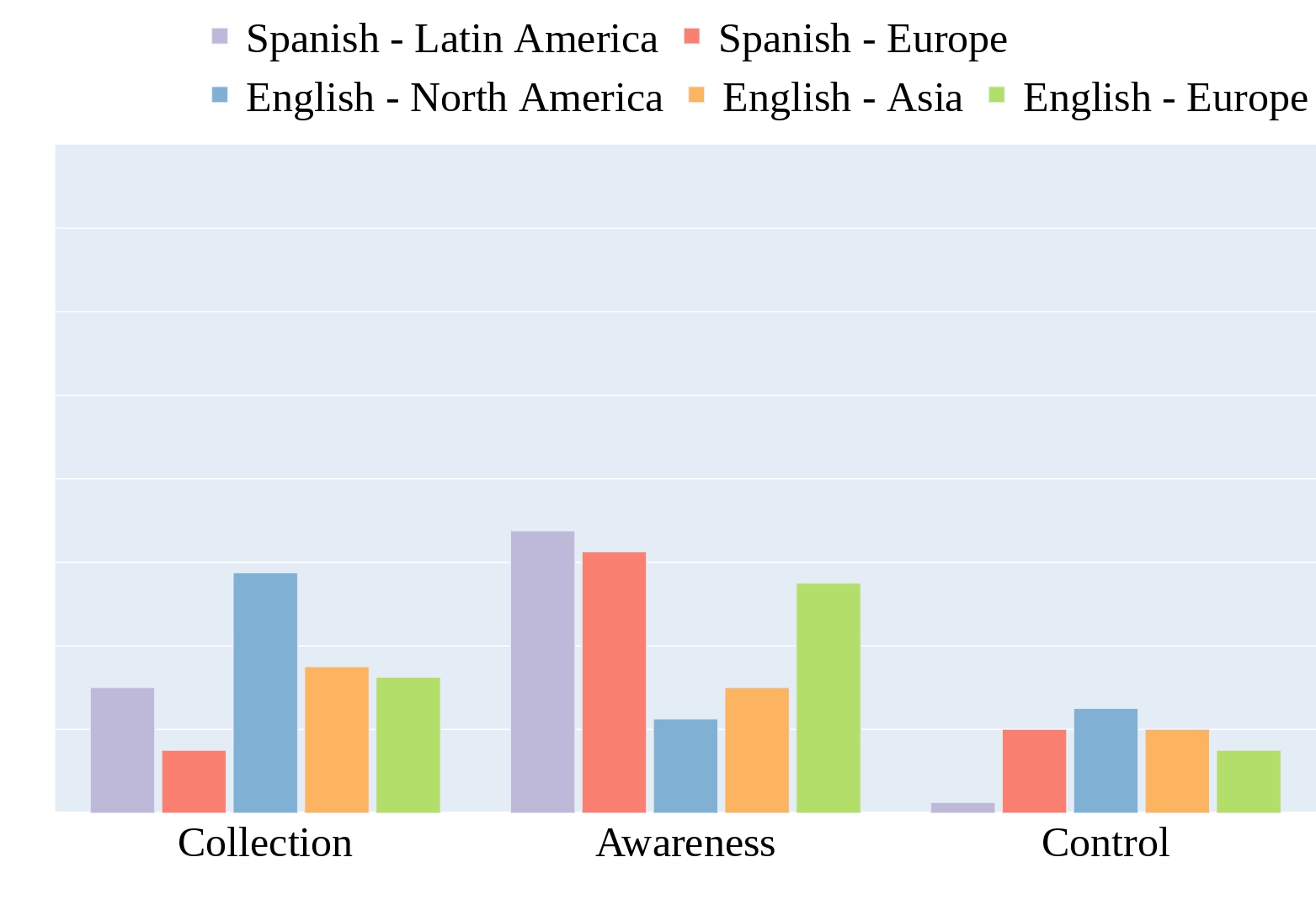} }}%
    \caption{Proportion of terms related to the IUIPC dimensions by language (a) and region (b)}%
    \label{fig:representation_iuipc_dimensions_by_language_region}
\end{figure}

\hlreview{We find no significant differences on the emphasis on information privacy concerns across languages or regions} (see Table {\ref{tab:null_hypothesis_IPC_tokens}). \hlreview{Thus, we cannot reject the null hypotheses. We conclude that IPC are present at similar rates in Spanish and English.} \hlreview{They cover a considerable proportion of the semantic contexts}
, with more than 30\% of terms in both languages. \hlreview{Considering the regional datasets, IPC describes between 20\% and 40\% of the terms. }
\celeste{While we observe some variation in emphasis on IPC across regions, with the largest proportion in the Latin American dataset and the smallest fraction in the Asian data, the differences across regions are not  enough to be statistically significant.}

\hlreview{The rest of the terms are better described by our initial categories, such as \textit{company}, \textit{information} and \textit{users}. Compared to the IPC category, all of them cover smaller fractions of the semantic contexts under study. }It should \hlreview{also} be noted that irrelevant categories (grouped as \textit{others}) add up to \hlreview{large proportions} in all datasets, ranging from 30\% to more than 60\%. \hlreview{Together, these results reveal that while a social media dataset about a data breach scandal does bring relevant content about information privacy concerns, this comes with a fair amount of noisy content. }

\begin{table}[!h]
\caption{
Results of Chi-squared tests to compare proportions of terms by language and world regions. The null hypothesis was rejected if $p<.0102$}
\label{tab:null_hypothesis_IPC_tokens}
\begin{tabular}{@{}p{0.67\textwidth}rrrr}
\toprule
\multicolumn{1}{c}{Null hypothesis} &
\multicolumn{1}{c}{$\chi^2$}&
\multicolumn{1}{c}{N} &
\multicolumn{1}{c}{DF} &
\multicolumn{1}{c}{$p$ value}
\\
\midrule
There is no difference in \% of \textit{IPC} terms between languages   & 0.15  & 110   & 1  & .70\\
There is no difference in \% of \textit{IPC} terms among world regions         & 8.04  & 237   & 4  & .09\\
\midrule
There is no difference in \% of \textit{collection} terms between languages   & 11.65 &31	&1&\textbf{\textless{}.001} \\
There is no difference in \% of \textit{collection} terms among world regions &	10.97	&68	&4	& .03 \\
There is no difference in \% of \textit{control} terms between languages	& 0.00&24&	1	&1.00\\
There is no difference in \% of \textit{control} terms among world regions	& 7.15	& 33	& 4	& .13 \\
There is no difference in \% of \textit{awareness} terms between languages  &	4.12	&41&	1&	.04
\\
There is no difference in \% of \textit{awareness} terms among world regions	&13.58&	95&	4	&\textbf{.009}\\
\midrule
There is no difference in \% of \textit{regulation} terms between languages & 0.69         & 13  & 1  & .41          \\
There is no difference in \% of \textit{regulation} terms among world regions             & 11.69        & 26  & 4  & .02      \\    
\bottomrule
\end{tabular}
\end{table}

\subsubsection{IUIPC dimensions}

Digging deeper \hlreview{in the terms related to information privacy concerns, }
we analyze the proportions of terms that match each IUIPC dimension across languages and world regions (see Figure {\ref{fig:representation_iuipc_dimensions_by_language_region}} and Table {\ref{tab:null_hypothesis_IPC_tokens}}). 

\hlreview{We observe a broader emphasis on
\textit{collection} in English} ($\chi^2(1, 31)=11.65, p=<.001$) \celeste{than in Spanish.} \nuevo{Cohen's effect size value (w = $.61$) suggests that this is  a high practical significance }\cite{cohen1988statistical}.  Even though this pattern seems to be influenced by a 
higher proportion on \textit{collection} 
in the English content from North America than in any other region} (Figure \ref{fig:representation_iuipc_dimensions_by_language_region}), \hlreview{regional differences are not statistically significant after multiple comparisons correction ($\chi^2(4, 68)=10.97, p=.03$).}

\hlreview{In turn, while we cannot reject a null hypothesis regarding differences on \textit{awareness} by language after corrections ($\chi^2(1, 41)=4.12, p=.04$), we find a significant difference across world regions ($\chi^2(4, 95)=13.58, p=.009$). }\nuevo{Cohen’s effect size value (w = .38) suggests a moderate to high practical significance.} \celeste{ Here, we calculated the standard residuals to determine which world regions make the greater contribution to this chi-square test result.  
We find that compared with other world regions, data in English from North America have a smaller ratio of awareness terms (chi-square standard residual = $-2.56$). The opposite is found in data in Spanish from Latin America (chi-square standard residual = $2.05$).  
}


\hlreview{Finally, we find no evidence to reject the null hypothesis regarding control.} Control is equally present in both languages and the regions under study.

\subsubsection{Regulation}

\begin{figure}[!h]%
    \centering
    \subfloat[\centering By Language]{{\includegraphics[width=0.49\linewidth]{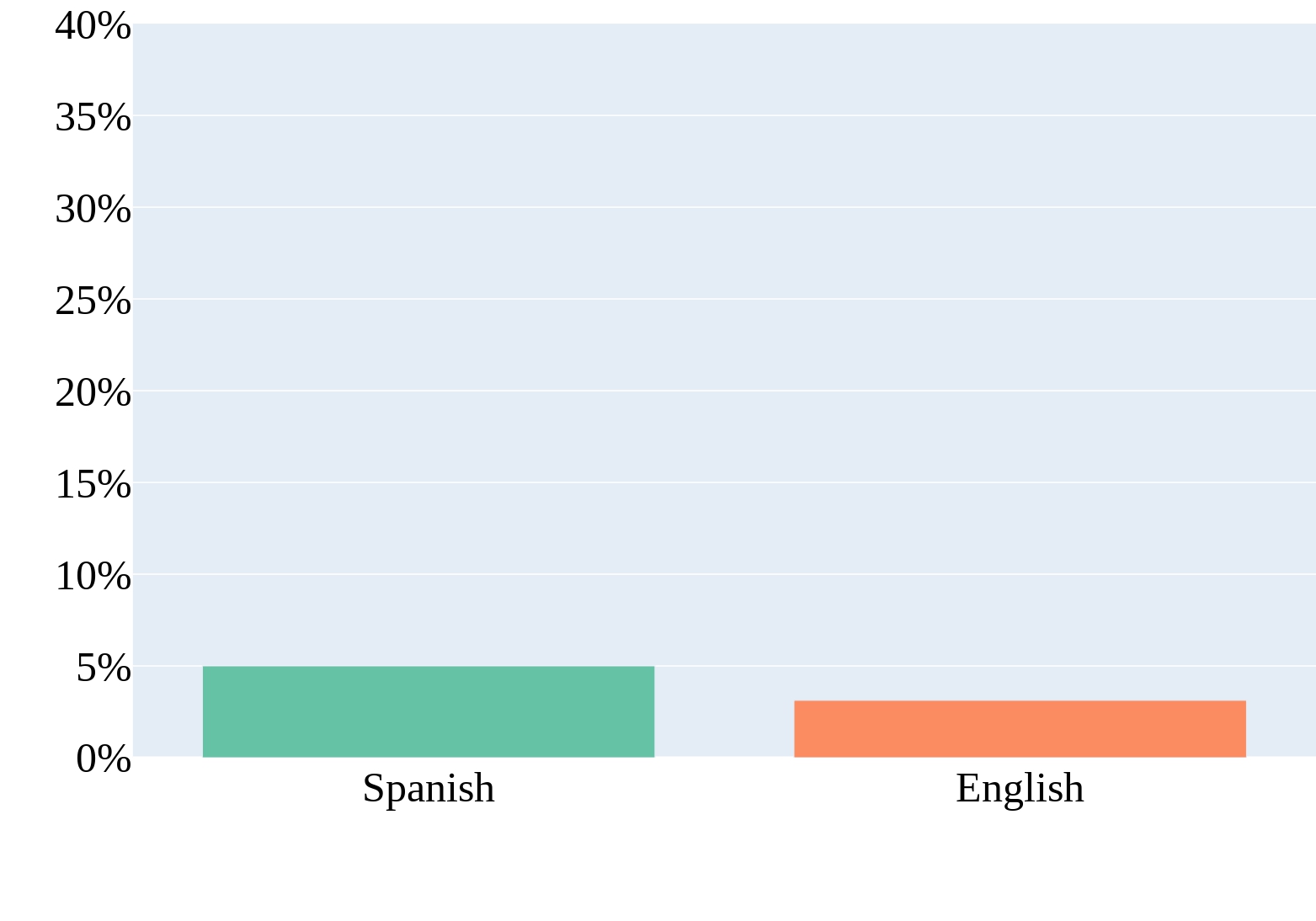} }}%
    \hfill
    \subfloat[\centering By world-region]{{\includegraphics[width=0.49\linewidth]{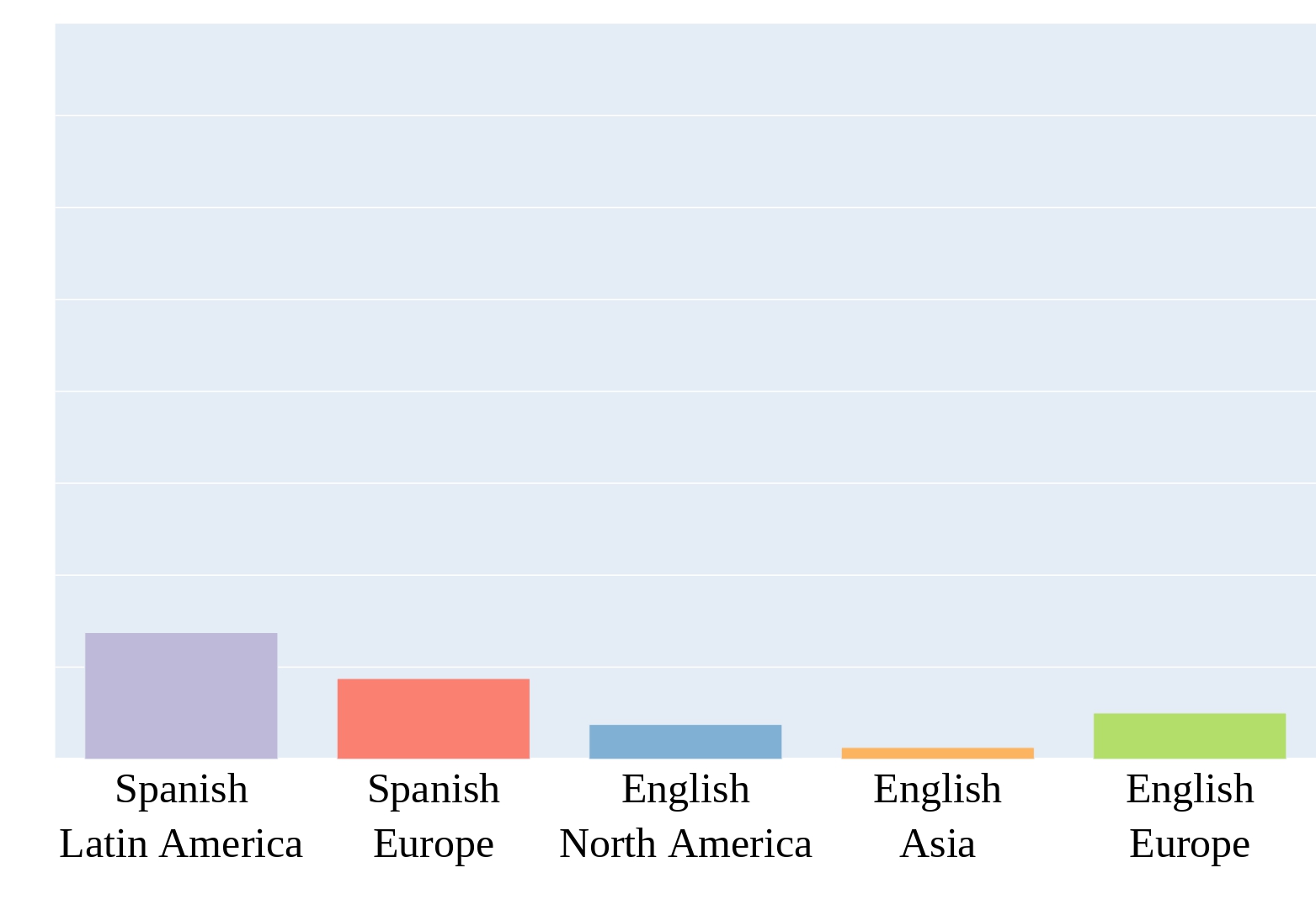} }}%
    \caption{Proportion of terms associated to regulations by language (a) and region (b)}%
    \label{fig:regulation_representation_by_language_region}
\end{figure}

Even though the concept of regulation is not part of the IUIPC dimensions, prior literature \cite{cockcroft2016relationship,lee2019information,da2018information} has suggested that it is related to people's concerns about information privacy. We find terms associated with this category in all our word embeddings (see Figure \ref{fig:regulation_representation_by_language_region}). \hlreview{However, the difference in proportions between Spanish and English data is not statistically significant ($\chi^2(1, 13)=0.69, p=.41$). Likewise, we do not find  enough evidence to reject the null hypothesis regarding differences across world regions after multiple comparison correction ($\chi^2(4, 26)=11.69, p=.02$) (see Table} {\ref{tab:null_hypothesis_IPC_tokens}}).

\section{Discussion}
\label{sec:discussion}
\hlreview{This work explores the potential of social media as a data source to study cross-language and cross-regional differences in information privacy concerns.} We conduct an analysis of \hlreview{Twitter} data related to a particular data breach \hlreview{news} to deepen our understanding of how \hlreview{people} from different world regions and who speak different languages frame privacy concerns. We chose to focus on the Cambridge Analytica scandal because it triggered a wide-ranging exchange on social media about user \hlreview{information and companies' data} practices. \hlreview{We build upon the potential of word embeddings to derive a semantic context of each term in a corpus. The contexts are built according to terms that are commonly used in the same phrases.}  By characterizing a keyword's nearby terms, we \hlreview{seek} to reveal the context in which a keyword was discussed \cite{rho2018fostering}. \hlreview{Based on more than a million non-duplicated, human-generated tweets, we generate word embeddings for data in Spanish and English and for data from Latin America, North America, Asia and Europe.} For each embedding, we conduct a qualitative analysis of 
the semantic contexts of four privacy-related keywords: \textit{information}, \textit{privacy}, \textit{company}, and \textit{users}. 
 
\hlreview{Collecting and analyzing the semantic contexts of these privacy-related keywords} 
allows us to observe \hlreview{the presence of terms related to} information privacy concerns in the collected tweets. Through iterative manual coding, we \hlreview{characterize the semantic contexts using 15 categories. Several of these categories are easily mapped to 
the three dimensions of the Internet User Information Privacy Concerns (IUIPC): \textit{collection}, \textit{awareness}, and \textit{control}} 
(See Figure \ref{fig:iuipc_dimensions_per_language_details}). In this way, we find evidence that social media content can reveal information about privacy concerns. 

Our approach \hlreview{takes} into consideration a vast amount of online \hlreview{content} posted freely and spontaneously on Twitter 
to create the semantic context of each keyword. Thus, it \hlreview{gives} a sense of a collective perspective on information privacy concerns by language and world region, which can become a complementary approach to current survey-based methods. \final{Our method aims to discover knowledge from a large-scale social media dataset in a topic for which a ground truth does not exist. Unfortunately, such ground truth is unlikely to exist because large-scale, multi-country, and multi-language surveys are too expensive to conduct} \cite{doi:10.1080/0144929X.2020.1831608}.  
\final{As an alternative approach, we used word embeddings to find the semantic contexts of relevant keywords and followed a qualitative approach to validate the results. We carefully analyzed more than a thousand terms of the semantic contexts, conducted open coding to formulate a data-grounded categorization, and then contrasted our categorization with IUIPC} \cite{malhotra2004internet}\nuevo{, one of the well-accepted theoretical conceptualizations of information privacy concerns}. \final{ While this is not the common ground truth of other natural language processing tasks such as classification, our process draws from qualitative approaches to validate the results of an automated text analysis. We discuss below how our findings extend our current understanding of privacy concerns and open new lines of inquiry.}

\hlreview{Beyond matching content to current conceptualizations of information privacy concerns, our results suggest a more granular categorization of one of them. Our results hint that \textit{awareness} might include more specific sub-topics that users can be aware of, such as \textit{privacy and security terms} (e.g., cybersecurity, confidentiality), \textit{security mechanisms} (e.g., credentials, encrypted), and \textit{privacy and security risks} (e.g., scams, grooming). The presence of terms that fit these categories reveals that they are already part of public online conversations around privacy. A distinction among broad privacy and security terms, mechanisms to protect data and potential data risks might be useful to further describe the kinds of knowledge people have. Additionally, awareness about some of these subtopics might be more influential than others. For example, knowing about risks and mechanisms might be a sign of higher privacy concerns while knowing broad privacy and security terms might not. The distinction between sub-topics could also guide users', educators' and practitioners' efforts to enhance information privacy literacy. Future work can explore the relevance of this distinction and its implications for information privacy practices.}

\hlreview{Besides, the presence of the \textit{regulation} category highlights its importance in relation to information privacy concerns. Regulation refers to laws or rules that aim to regulate the use of personal data. They have often been considered a factor influencing information privacy concerns}\cite{ebert2020does,benamati2021information}\hlreview{. The emergence of this category from our open coding confirms its relevance through its frequent appearance in public posts about a data breach scandal. Such relevance might be related to the elaboration of laws and public policies about data usage worldwide. These regulations are not only a topic of data and law experts, but it seems to be part of the public discourse around data privacy online. It is noticeable the common presence of a specific regulation, the GDPR, in our datasets. } \celeste{GDPR is a privacy regulation that has been in effect since late May 2018 in the European Union \mbox{\cite{gellman2019fair}}. Our data collection period covered the early months of its implementation. This regulation prohibits processing and  exploiting personal data such as health status, political orientation, sexual preferences, religious beliefs, and ethnic origin. Thus, it aims to decrease the privacy risks that may derive from malicious use of such information, including cases like the Cambridge Analytica scandal \mbox{\cite{cabanas2018facebook}}. GDPR seeks to convert individuals into empowered citizens involved in the decision-making process related to their personal information \mbox{\cite{karampela2019exploring}}. As an example, with this regulation in effect, companies are required to inform individuals about their rights (e.g., restriction of processing, erasure of data), the storage period of data, and additional sources that have been used to acquire personal data \mbox{\cite{ebert2020does}}}. 
\hlreview{The explicit presence of GDPR in our data
might be evidence of its influence on shaping people's arguments about privacy concerns and its importance not only in Europe but worldwide. Further work can focus on exploring how to integrate better the role of regulations into the current conceptualizations of information privacy concerns, which were proposed long before data privacy regulations were as common as they are now around the globe. Moreover, future work could also explore the interaction between regulations and specific information privacy concerns dimensions.} 

\hlreview{While we find similar rates of terms related to information privacy concerns across languages and regions, we observe significant differences in emphasis on collection and awareness.} These results indicate that different groups view the Cambridge Analytica scandal from a particular standpoint. It is important to notice though that while information privacy terms appear through our method, they also come along with a considerable amount of other terms that we consider noisy data. 
Nevertheless, our findings show the potential \hlreview{of using social media data} for cross-language and cross-regional comparisons to identify similarities and nuanced differences on privacy-related perspectives worldwide. 

Our analysis reveals that the semantic contexts generated by tweets \hlreview{written in English have significantly} more terms related to \textit{collection} than those \hlreview{written in Spanish}. This is a novel finding. When freely expressing online about privacy keywords, English speakers give significantly more emphasis to data collection than Spanish speakers. This difference can lead researchers and practitioners to explore the effectiveness of more tailored data privacy campaigns to specific populations. For example, populations that are more concerned about collection might need more information about the benefits of sharing their information to be able to make a decision about it. A high emphasis on collection in English is also congruent with prior literature observing that college students from the USA are more worried about collection of personal information than control over it \cite{yang2013young}. Exploring if this trend is shared by people from other English speaking countries can help clarifying which of these patterns are better explained by location or language.

Future work can explore why we observe a significant language difference in emphasis on collection. A feasible explanation might be related to the users' country of residence. \hlreview{Note that our tweets in English come mainly from the USA and UK.} \hlreview{Both were the countries most closely connected to the Cambridge Analytica scandal due to the misuse of data for political campaigns in the USA's 2016 presidential election and Brexit \mbox{\cite{cadwalladr2018revealed}}. It is possible that this shared experience resulted in a larger emphasis on collection in the English than the Spanish data. An alternative hypothesis is associated with differences in regulations.} Information privacy concerns \hlreview{might be} a reflection of customer privacy regulations in their respective countries \cite{markos2017information,kumar2018customer}. In contrast to European countries, \hlreview{that have} 
adopted a data protection directive from a \textit{government-imposed} perspective, the USA has followed an \textit{industry self-regulation} \mbox{\cite{kumar2018customer}}. Considering that companies have more freedom to collect and process personal data in \hlreview{North America}, 
it would be reasonable that data collection practices are of deeper concern to individuals from North America than those in Europe.  
\hlreview{This could also be supported by our data when observing that North America has the highest proportion of terms related to \textit{company} (see Figure }\ref{fig:global_categories_by_language_region}), which we also found in our prior work using a different text mining method and a smaller dataset \cite{gonzalez2019global}. \hlreview{However, our data analysis does not support the hypothesis of regional differences. It is possible that our data does not have enough power given the multiple comparisons we conducted. Future research is needed to explore alternative hypothesis that can explain the broader emphasis on collection among English speakers, compared to Spanish speakers. } 

\hlreview{We also observe significant regional differences on \textit{awareness}. Particularly, data from North America shows the smallest emphasis on \textit{awareness} while Latin America has the highest. Given that most studies on information privacy concerns are centered on the USA, this finding is particularly important. It warns us against the (sometimes implicit) assumption that North American data about privacy concerns can be generalizable to other regions. At least regarding emphasis on awareness, we find evidence that data from the USA is not similar to other regions. Thus, this result provides observational evidence to argue that it is necessary to include more diverse populations to have more a accurate understanding of the phenomena around data privacy. This finding also invites practitioners to address other regions, such as Latin America, using more different approaches in their terms of services and privacy policies. Populations that are more concerned about awareness might be more receptive to companies that use more transparent communications of their use of personal data, for example.}

\hlreview{It is worth noting that Latin American shows the largest emphasis on \textit{awareness}.} \hlreview{Our results provide evidence of a disconnection between Latin America and North America regarding this aspect. It is possible that this broad interest on awareness can be a reflection of a connection of Latin America to the European perspective on data privacy.} Latin America presents a \hlreview{particular} scenario. It lies between two different approaches to personal data regulation: the principles contained by the European GDPR and the fragmented framework of the USA, where data protection is divided by sector \cite{aguerre2019digital}. Privacy regulations are considered an essential concern for many Latin American countries, and after data privacy breaches such as the Cambridge Analytica one, this issue has received increased attention in the public opinion and policy spheres in the region \cite{aguerre2019digital}. Previously, researchers have argued that GDPR could be one of the most influential pieces of data protection legislation ever enacted with influence beyond Europe \cite{kuner2017gdpr}. Indeed, in Brazil, a new GDPR-like law (\textit{Lei Geral de Proteção de Dados Pessoais, LGPD, in Portuguese}) \hlreview{has} become effective since August 2020 \cite{dias2020perceptions}. \hlreview{Future studies can explore connections among data privacy regulations worldwide, how they relate to public opinion on the issues of privacy, and how they are influenced by national and international data breaches. }



\revisado{As we found regional but not language differences in emphasis on privacy concerns, we conducted a follow-up analysis to assess whether there is a language difference within a single region. Europe was the only region where we had enough data in both languages to conduct such a comparison. We did not find significant differences in emphasis of any IUIPC dimension between data in English and Spanish from Europe ($\chi^2(3, 92)=0.15, p=.98$). There was no evidence of significant differences in emphasis on regulations either. Thus, this additional analysis provides additional evidence to support that information privacy concerns are more related to the region of residence than the spoken language. Nevertheless, further research is required to understand better the role of regulatory regimes, consumer practices, and economic development factors on these differences \mbox{\cite{OKAZAKI2020458}}. As the Spanish-English balance in tweets in our dataset is such that it does not lend itself to intra-region comparison for Asia and North and Latin America, future work could seek to explore if this pattern repeats in those regions as well}


As with any study, our research has limitations. We collected data through the free standard streaming Twitter API using specific hashtags and keywords. Thus, we only had access to a limited sample of all the tweets about the scandal. \hlreview{We used Botometer to detect and remove tweets likely to be created by bots. This tool can only analyze Twitter public accounts; therefore, it could not be used on suspended accounts or those with their tweets protected when running our analysis. We decided to remove these accounts' tweets from our datasets because we can not confidently claim that humans generated them. Indeed, previous research suggests that it is likely that social bots were present in this cohort \mbox{\cite{8537833}}. 
}
\hlreview{Moreover, we focused our investigation on four keywords in English: \textit{information}, \textit{privacy}, \textit{users}, and \textit{company} and their corresponding translations to Spanish. While using synonyms would have brought similar semantic contexts, adding more concepts can strengthen the results. Future work can explore 
other keywords 
such as: 
\textit{intimacy}, 
and \textit{consumers}. } \revisado{Similarly, we did not use the terms \textit{user}, and \textit{companies} as keywords. While word embeddings capture syntactic regularities such as singular/plural forms \mbox{\cite{mikolov2013distributed,9259855}}, we reason that this methodological decision should not have affected considerably our results. Nevertheless, future work could include plural and singular versions of the same term to confirm this hypothesis.} \hlreview{The sample size of our manual coding process (40 words per keyword in each embedding) could have impacted the results. We chose the number of retrieved terms after manually inspecting the list of nearest words by each keyword in all our embeddings. We picked a threshold that allowed us to obtain a high number of meaningful words in most embeddings. Higher thresholds make it more likely to include terms with no apparent relation to the keywords (e.g., v.i.p; ground; approximately). In word embeddings with reduced vocabularies like ours, the number of relevant terms available for a specific keyword is limited. This characteristic explains why the number of irrelevant terms (\textit{Other} in Figure 3) is high in datasets with small vocabularies, such as the Spanish and English-Asia datasets. Future work could evaluate how sensitive our approach is to changes in vocabulary size and threshold for the nearest terms. This decision may introduce a bias in the results, and it is one of the limitations of our approach of social media textual data.}
%

\section{Conclusion}
\label{sec:conclusion}


We 
\hlreview{conducted} a cross-language and cross-regional study on social media content about a major data privacy leakage: the Cambridge Analytica scandal. We categorized our Twitter data into two different languages and four geographical regions. 
Our results shed light on \hlreview{language and} regional differences on information privacy concerns by 1) creating word embeddings by language and world regions to leverage social media data about a data breach scandal, 2) conducting open coding and content analysis of the semantic contexts (generated by the embeddings) of privacy-related keywords, 3) mapping the results to a well-known information privacy framework, and (4) conducting a comparative analysis across two languages and four world regions. 

We found that data \hlreview{in English} 
shows a broader emphasis on data collection
, \hlreview{while data from North America shows the smallest emphasis on awareness. In turn, data from Latin America has the broadest emphasis on awareness.} We discuss how our findings extend current conceptualizations of information privacy concerns, and how they might relate to regulations about personal data usage in the regions we analyzed.  

Future work can dig deeper on the differences we observed and explore further the potential causes we discussed. Future studies might build upon our work to examine privacy concerns considering more languages, more geographical locations \hlreview{or different information privacy frameworks.} Using our methodology to compare datasets across longer periods of time could be useful to determine if the semantic contexts of the privacy keywords changes over time.

\section{Acknowledgments}
\hlreview{The authors  want to thank Francisco Tobar, MSc. Computer Science student at Universidad Técnica Federico Santa María, for helping us to strengthen our findings through statistical analysis.} \nuevo{ Moreover, we acknowledge anonymous reviewers for insightful comments that helped
us revise and refine the paper. } 

\section{Funding and conflicts of interests}
This collaboration was possible thanks to the support of the Fulbright Program, under a 2017-18 Fulbright Fellowship award. This work was also partially funded by CONICYT Chile, under grant Conicyt-Fondecyt Iniciaci\'on 11161026. The first author acknowledges the support of the PIIC program from Universidad T\'{e}cnica Federico Santa Mar\'{i}a and CONICYT-PFCHA/Mag\'{i}sterNacional/2019-22190332. The authors declare that there is no conflict of interest regarding the publication of this paper.
\bibliographystyle{cscwjournal}
\bibliography{new_references}

\end{document}